\documentclass[fleqn,showkeys,reprint]{revtex4-1}
\usepackage{graphicx,epsfig}
\usepackage{subscript}
\usepackage{amsmath,amssymb,amsfonts}
\usepackage{xcolor}

\begin{document}
 \title{Delay time and Non-Adiabatic Calibration of the Attoclock
\\ \it Multiphoton process versus tunneling in strong field interaction}
\author{Ossama Kullie}
\affiliation{Theoretical Physics, Institute for Physics,
 Department of Mathematics and Natural Science, Universit{\"a}t Kassel, 
 34132 Kassel, Germany}
\email{kullie@uni-kassel.de}
 \author{Igor Ivanov}
 \affiliation{Center for Relativistic Laser Science, Institute for
  Basic\ Science (IBS), Gwangju 61005, Republic of Korea}
\begin{abstract}
\scriptsize
 The measurement of the tunneling time in attosecond experiments, 
 termed attoclock,
 triggered a hot debate about the tunneling time, the role of time in 
 quantum mechanics, where the interaction with the laser pulse involves 
 two regimes of a different character, the multiphoton and the 
 tunneling (field-) ionization. 
 In the adiabatic field calibration, one of us (O. K.) developed in 
 earlier works a real tunneling time model and showed that the model 
 fits well to the experimental data of Landsmann et al. (Optica 
 {\bf 1}, 343 2014).  
 In the present work, it is shown that the model explains the 
 experimental result in the nonadiabatic field calibration, where one
 reaches a good agreement with the experimental data of Hofmann et al. 
 (J. of Mod. Opt. {\bf 66}, 1052, 2019). 
 Furthermore, we confirm the result with the numerical integration 
 of the time-dependent Schr\"odinger equation.
 The model is appealing because it offers a clear picture of the 
 multiphoton and tunneling field-ionization regimes.
 In the nonadiabatic case (the nonadiabatic field calibration), the 
 ionization is mainly driven by multiphoton absorption.
 Surprisingly, at a field strength $F \le F_a$ ($F_a$ is the atomic field 
 strength) the model always predicts a time delay with respect to 
 the quantum limit $\tau_a$ at $F=F_a$.
 For an adiabatic tunneling,  
 the saturation at the limit ($F=F_a$) explains the well-known Hartman 
 effect or Hartman paradox. 
 \end{abstract} 
\maketitle
\footnotesize
\section{Introduction}\label{sec:int}
 The measurement of the tunneling time, in the strong field laser-matter 
 interaction and attosecond science, triggered a hot debate about 
 the tunneling with its adiabatic and nonadiabatic frontiers, the 
 multiphoton and the intermediate regimes, the tunneling time and the 
 role of time in quantum mechanic.
 Tunneling happens when the interacting electron is field-ionized 
 by a tunneling mechanism, which can occur when the field strength of 
 the laser pulse is strong enough but smaller than the atomic field 
 strength $F_a$ of the system (an atom or a molecule). 
 $F_a$ is defined by the ionization potential of the valence or the 
 interacting electron. 
 In the adiabatic tunneling, Kullie presented in a previous work 
 \cite{Kullie:2015} a tunneling model, in which the tunneling time 
 (T-time) is a time delay with respect to ionization time at atomic 
 field strength $F_a$. 
 The tunneling time delay showed a good agreement with the 
 attoclock result or the attosecond (angular streaking) experiment of 
 Landsmann et al. \cite{Landsman:2014II,Eckle:2008s,Eckle:2008} for 
 Helium (He-) atom \cite{Kullie:2015}, and of Sainadh et al.  
 \cite{Sainadh:2019} for Hydrogen (H-) atom \cite{Kullie:20181} (apart 
 from a factor $1/2$) and with the accompanying numerical integration of  
 the Schr\"odinger equation (NITDSE) of \cite{Sainadh:2019}. 
 Furthermore, the T-time picture of \cite{Kullie:2015} shows an 
 intriguing similarity to the famous Bohr-Einstein weighing {\it photon  
 box Gedanken experiment (BE-photon-box-GE)} \cite{Aharonov:2000}, 
 \cite{Auletta:2009} (p. 132), where the former can be considered as 
 a realization of the latter. 

The description of the tunneling ionization process which we propose 
differs from the traditional approach based on the evaluation of the 
expression for the ionization amplitude obtained by using the saddle 
point method (SPM). 
Such an approach leads ultimately to the well-known imaginary time (ITM) 
\cite{Popov:2005,Perelomov:1966} and quantum orbit (QO) 
\cite{Popruzhenko:2014,Lewenstein:1994} methods, which offer the following 
interesting and appealing picture of the ionization process.
An electron trajectory (generally complex) originates at the (complex) 
moment of time corresponding to the saddle point,
descends further on the real time-axis, and propagates in the real time 
after intercepting the real time-axis. The real part of the complex saddle 
point can be interpreted as the moment when the electron begins the 
under-the-barrier motion (hence the complex-valued time and velocity),
and the point of intercept of the quantum trajectory with the real 
time-axis as the moment of time when the electron exits the tunneling 
barrier or the time when the ionization event occurs. 
This picture proved to be extremely useful in understanding the tunneling 
ionization process. We should note, however, that it is no means unique 
description of the tunneling ionization. 
Indeed, the description based on the SPM and complex quantum 
trajectories is but convenient way to visualize ionization process. 
This can be seen already from the fact, that the quantum trajectory 
used for the evaluation of the action in the quantum orbits method is 
not unique. Due to the analycity properties of the integrand defining 
the action it can be deformed as long as the integration path does not 
cross any singular points, which makes the location of the point where 
the quantum orbit intercepts the real time axis (and hence the very 
notion of the barrier exit time) somewhat arbitrary 
\cite{Popruzhenko:2014}.  
Furthermore, the very use of the SPM is based on the assumption that 
the vector potential describing the electric field is an analytic (in the 
sense of the theory of analytic functions of a complex variable) 
function of time. 
That need not be necessarily the case, one can easily imagine 
a function of time describing a perfectly realistic pulse shape which 
is not a complex analytic function of time. 
An alternative description of the tunneling ionization process must, 
therefore, exist, which does not make use of the complex ionization 
time and complex electron trajectories. 
We describe such a description in the present work.

\begin{figure}[t]
 \vspace{-6.20cm}
 \resizebox{15cm}{!}{\includegraphics{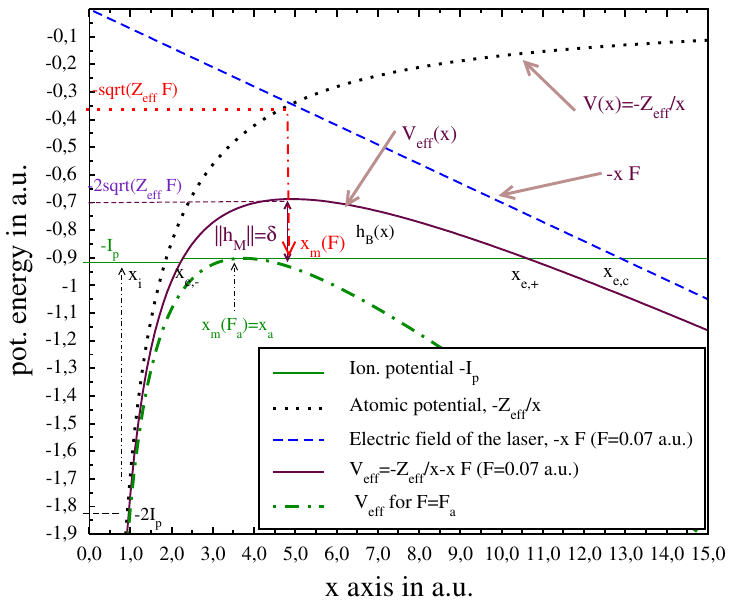}}
 \vspace{-8.0cm}
 \caption{\label{Hofdata1}\footnotesize  (Color online)
 Graphic display of the potential and the effective potential curves, the two inner 
 and outer points $x_{e,\pm}={(I_p\pm\delta_z)/2F}$, 
 $\delta_z=\sqrt{Ip^2-4 Z_{eff} F}$, the barrier width 
 $d_B=x_{e,+}-x_{e,-}=\delta_z/F$. 
 $I_p$ is the ionization potential, $Z_{eff}$ is the effective nuclear charge 
 and $F$ is the electric field strength of the 
 laser pulse at maximum. $x_{e,c}=I_p/F\equiv d_C$ is the ``classical-'' exit point and 
 $x_m(F)=\sqrt{Z_{{eff}/F}}$ is the 
 position at the maximum of the barrier height $h_B(x)$, and $x_a=x_m(F=F_a)$, $F_a$ 
 is the atomic field strength,  
 see text. The plot is for He-atom in the SAEA with $Z_{eff}=1.6875$ and $I_p=0.90357\, au$. 
 For systems with 
 a different $Z_{eff},\, I_p$ the overall picture stays the same.}
 \end{figure}

 In our model, introduced in \cite{Kullie:2015} (see Fig. 1), an 
 electron can tunnel and is field-ionized by a laser pulse 
 with a peak (electric) field strength $F$. 
 A direct ionization (no tunneling) happens when the field strength 
 reaches a threshold value called atomic field strength $F_a=
 I_p^{2}/(4 Z_{eff})$ \cite{Augst:1989,Augst:1991}, where $I_p$ is the 
 ionization potential of the system (atom or molecule) and $Z_{eff}$ is 
 the effective nuclear charge in the single-active-electron 
 approximation (SAEA). 
 For $F<F_a$, ionization can happen by tunneling through a barrier  
 which is  built by an effective potential due to the Coulomb 
 potential of the nucleus and the electric field of the laser pulse. 
 It can be expressed in the length gauge (due to G\"oppert-Mayer 
 gauge-transformation \cite{Goeppert:1931}) in a one-dimensional form 
 \begin{equation}\label{vef}
  V_{eff}(x)=V(x)-x F =-\frac{Z_{eff}}{x}-x F,
 \end{equation}
 compare Fig. \ref{Hofdata1}. 
 In eq \ref{vef} and hereafter, we adopt the atomic units ($au$), 
 where the Planck constant, the electron's mass and charge are set 
 to unity, $\hbar = m = e = 1$.
 In the model the tunneling process can be described solely by 
 the ionization potential $I_p$ of the valence (the interacting) 
 electron and the peak field strength $F$, which leads to the quantity   
 $\delta_z=\sqrt{I_p^{2}-4 Z_{eff} F}$, where $F$ stands (throughout 
 this work) for {\it the peak electric field strength at maximum}. 
 In Fig.  \ref{Hofdata1} (for details see \cite{Kullie:2015}), the inner 
 (entrance $x_{e,-}$) and outer (exit $x_{e,+}$) points are given by   
 $x_{e,\pm} =(I_p\pm\delta_z)/(2F)$, the barrier width $d_B=
 x_{e,+}-x_{e,-}=\delta_z/F$, and the barrier height 
 (at $x_m(F)=\sqrt{Z_{eff}/F}$) is $h_B(x_m)=I_p-V_{eff}(x_m)=\delta_z$.
 Which can also be obtained by two real quantities (indicating a symmetry), $h_M^{\pm}(x_m)
 =(-I_p\pm\sqrt{4 Z_{eff}F})$  
 and we have  $\|h_M\|=|h_M^{+}h_M^{-}|^{1/2}=\delta_z\equiv\overline{h_B}$.  
 At $F=F_a$ we have $\delta_z=0$ ($d_B=\overline{h_B}=0$), the barrier  
 disappears and the direct or the barrier-suppression ionization (BSI)
 starts \cite{Delone:1998}. 

 In the (low-frequency) attosecond experiments, the laser field is 
 comparable in strength to the electric field of the atom. 
 Usually, one uses intensities of the order of $10^{14}\, W cm^{-2}$. 
 
 A key quantity is the Keldysh parameter \cite{Keldysh:1964},
 \begin{equation}\label{gamK}
 \gamma_{_K} = \frac{\sqrt{2I_p}}{F}\, \omega_0=\tau_{K}\, \omega_0, 
 \end{equation} 
 where $\omega_0$ is the central circular frequency of the laser pulse  
 and $\tau_{K}$ denotes the Keldysh time.  
 
 According to Keldysh or strong field approximation (SFA), for values 
 $\gamma_{_K}>1$ (actually $\gamma_{_K}\gg 1$), the dominant process 
 is multiphoton ionization (MPI). 
 In the opposite case, for $\gamma_{_K}<1$ (actually 
 $\gamma_{_K}\ll 1$), a field-ionization can happen by a tunneling 
 process, which occurs for $F<F_{a}$.  
 This picture has been subsequently refined and is known now as the 
 Keldysh-Faisal-Reiss (KFR) theory \cite{Faisal:1973,Reiss:1980}.
 
 As we will see, this separation between tunneling and multiphoton 
 regime by Keldysh parameter $\gamma_K$ is not rigorous, since it 
 points to two case limits. 
 In our model, as we will see, the classification of the two regimes 
 (tunneling and multiphoton) is presented more clearly, when the 
 nonadiabatic ionization (nonadiabatic field calibration of Hofmann 
 et al.  \cite{Hofmann:2019}) is considered.  
 Like the adiabatic case, considered in \cite{Kullie:2015}, in the 
 present work, we find that in the nonadiabatic field calibration, the 
 field-ionization time is a time delay with respect to the quantum limit 
 at atomic field strength $F_{a}$.  

 In the experiment with He-atom \cite{Landsman:2014II,Hofmann:2019}, an 
 elliptically polarized laser pulse is used with $\omega_0=0.061991\,au\, 
 (\nu\approx 735\,nm)$, with ellipticity $\epsilon=0.87$. 
 The calibrated electric field strength is in the range $F\approx 
 0.02-0.10$ in the nonadiabatic ($F\approx0.04-0.11$ in the adiabatic) 
 case and for He-atom $I_p=0.90357\, au$.

In the attosecond angular streaking experiment, one uses {\it 
a close-to-circular polarized laser pulse, thereby ensuring a unique  
relationship between the time at which the electron exits the tunnel 
(the potential barrier) and the direction of its momentum after the 
laser pulse. 
The measured momentum vector of the electron hence serves as the hand 
of a clock, indicating the time when the electron appeared from 
the tunnel in the laser field, see \cite{Eckle:2008s}. }
 The main result of the tunneling model of Kullie \cite{Kullie:2015},  
 can be summarized with the following T-time formulas,
 \begin{eqnarray}\label{Tdi}
 \tau_{_{T,d}}=\frac{1}{2(Ip-\delta_z)}, \quad 
 \tau_{_{T,i}}=\frac{1}{2(I_p+\delta_z)}, \quad \\\nonumber
 \tau_{tot}=\tau_{_{T,i}}+ \tau_{_{T,d}}=\frac{I_p}{4Z_{eff} F}
 \end{eqnarray}
 The physical meaning of the relations is the following:
 the presence of a barrier causes a time delay $\tau_{_{T,d}}$, 
 which is a time delay with respect to the ionization at atomic field  
 strength $F_a$, as it is defined (from the appearance intensity) by 
 August et al. \cite{Augst:1989} (see \cite{Kullie:2015} and the 
 references therein), when the barrier disappears, $\delta_z(F_a)=0, 
 d_B(F_a)=0$.
 It is the time interval during which the electron is tunnel-ionized, 
 i.e. it ``passes'' the (emerging) barrier region (though semi-classically 
 determined) and escapes at the exit point $x_{e,+}$ to the continuum 
 \cite{Kullie:2015}.
 Whereas $\tau_{_{T,i}}$  is the time it takes to reach the entrance  
 point $x_{_{e,-}}$ from its initial position close to $x_i$, compare Fig. 
 \ref{Hofdata1}. 

 At the limit $F\rightarrow F_a\, (\delta_z\rightarrow 0)$ the two 
 steps coincide, the total time is $\tau_{tot}=\frac{1}{Ip}$ or 
 $\tau_{_{T,d}}=\tau_{_{T,i}}=\frac{1}{2Ip}$.
 For $F> F_a$ we enter the BSI regime \cite{Delone:1998,Kiyan:1991}, 
 which is outside the scope of the present work.
 At the opposite limit, we have $F\rightarrow 0$,
 $\delta_z\rightarrow I_p$ and $\tau_{_{T,d}}\rightarrow \infty$. 
 Hence, nothing happens, and the electron remains undisturbed in its 
 ground state, which shows that our model is consistent. 
 For details, see 
 \cite{Kullie:2015,Kullie:2016,Kullie:2018,Kullie:2020}. 

\section{Tunneling and Field-Ionization time delay}\label{sec:tt}
 In this section, we show that the time delay in our tunneling model 
 eq \ref{Tdi} can be understood differently, which explains the  
 nonadiabatic effects in principle through a multiphoton absorption, as 
 far as the nonadiabatic field calibration is considered, as done by 
 Hofmann et al. \cite{Hofmann:2019}. 
 eq \ref{Tdi} can be decomposed in a twofold time delay with respect to 
 ionization at $F_a$. 
 It explains the T-time in a more advanced picture.
 We can rewrite the T-time $\tau_{T,d}$ in eq \ref{Tdi} as follows
\begin{eqnarray}\label{TdF}
\tau_{_{T,d}}&=&\frac{1}{2(Ip-\delta_z)}=  
\frac{1}{2}\frac{I_p}{4Z_{eff}F} 
 \left(1+\frac{\delta_z}{I_p}\right)\\\nonumber 
 &=&\frac{1}{2I_p}\, \frac{F_a}{F}
 \left(1+\frac{\delta_z}{I_p}\right)\equiv\tau_a\, \chi(F)\\\nonumber
 &=&\frac{1}{2I_p}\frac{F_a}{F}
 +\frac{1}{2I_p}\frac{F_a}{F} \, \frac{\delta_z}{I_p}
 \equiv\tau_a \zeta_{F} + \tau_a \Lambda_{F}\\\nonumber
 &=&\tau_{dion}+\tau_{delt}, 
\end{eqnarray}
 where for clarity we define the dimensionless enhancement functions 
 $\chi(F),\,\zeta_{F},\,\Lambda_{F}$ for the quantum mechanical time limit 
 $\tau_a$. 
 The second line in eq \ref{TdF} immediately shows that our tunneling 
 time can be easily interpreted as a time delay with respect to 
 ionization time at atomic field strengths 
 $\tau_{_{T,d}}(F_a)=\tau_a=1/(2I_p)$, which is real and quantum 
 mechanically does not vanish.
 $\chi(F)$ is an enhancement factor for field strength $F<F_a$ (an 
 similarly $\zeta_{F},\,\Lambda_{F}$). 
 In the third line, we see that both terms are real and indicate 
 time delays. 
 The second term, $\tau_{delt}$, is real because $\delta_z>0$  
 is a real quantity \cite{Kullie:2015}. 
 \begin{figure}[t]
 \resizebox{8.25cm}{!}{\includegraphics{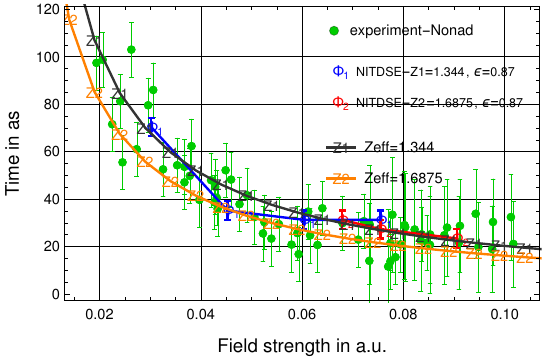}}
 \vspace{-0.30cm}
 \caption{\label{Hofdata2}\footnotesize (Color online)
 Graphic display of time delay versus field strength for He atom. 
 The time delay 
  $\tau_{dion} (\tau_{sym})$ as given in eq \ref{dion} 
 (eq \ref{Tsym}) for two $Z_{eff}$ values, $1.344 =\sqrt{2 Ip}\,(Z_1)$, 
 $1.6875\, (Z_2)$ \cite{Clementi:1963}, where $I_p$   
 is the ionization potential of He-atom. 
 Together with he experimental data of Hofmann in the new calibration 
 of the field strength \cite{Hofmann:2019}. 
 The NITDSE (see sec. \ref{sec:apdx} and \cite{IAIvanov:2014}:  
 $\Phi_1$ (blue) $Z_{eff}=1.344$ and $\Phi_2$ (red)   
 $Z_{eff}=1.6875$), with error bars of one degree ($\sim 4\, as$).} 
 \end{figure}
 However, as we will see, our picture corresponds to the imaginary  
 T-time picture in the case of the adiabatic field calibration 
 \cite{Kullie:20181,Kullie:2020}.
 In our twofold time delay picture, the first term is 
 a field-ionization time delay solely because $F$ is smaller than the 
 atomic field strength $F_a$, whereas the second term is a time delay 
 due to the barrier itself, which is the actual T-time as discussed in 
 detail in the recent work \cite{Kullie:2020}, due to the presence of 
 the factor $(\frac{\delta_z}{I_p})$ that specifies the (emerging) 
 barrier height $\delta_z$ relative to the initial one $I_p$, in 
 addition to the the factor $({F_a}/{F})$, further below in sec. \ref{sec:dis}.
 
 We note that the separation in a twofold time delay in eq \ref{TdF}, 
 represents a unified T-time picture in accordance with the Winful 
 UTTP \cite{Winful:2003} for the quantum tunneling of a wave packet or 
 a flux of particles scattering on a potential barrier.  
 According to Winful \cite{Winful:2003} the group time delay or the 
 Wigner time delay can be written in the form 
 \begin{equation}\label{Winf}
  \tau_g=\tau_{si}+\tau_{dwel},
 \end{equation}
 where $\tau_{dwel}$ is the well-known dwell time, see for example 
 \cite{Landsman:2015}, which can be seen to be corresponds to our 
 $\tau_{delt}$, and $\tau_{si}$ is, according to Winful, 
 a self-interference term, which corresponds to our $\tau_{dion}$ 
 in eq \ref{TdF}.  
 
 From  eq \ref{TdF}, we see that the two parts categorize two time 
 delays with respect to the atomic field strength.
 An ionization $\tau_{dion}$ (with no tunneling contribution) and 
 (an actual) tunneling $\tau_{delt}$, where the disruption by the 
 laser-field triggers the entire process. 
 That is different from (but it does not contradict) the well-known 
 premises of the strong field ionization theory, where  one commonly 
 uses $\gamma_K$ to divide the process into two regimes, the 
 multiphoton $\gamma_K\gg1$ and the tunneling $\gamma_K\ll 1$ regime. 
 In addition, the Keldysh time $\tau_K$ of eq \ref{gamK} is rather 
 a classical quantity, see below sec. \ref{sec:dis}. 
 \begin{figure}[t]
 \vspace{-0.65cm}
 \resizebox{8.25cm}{!}{\includegraphics{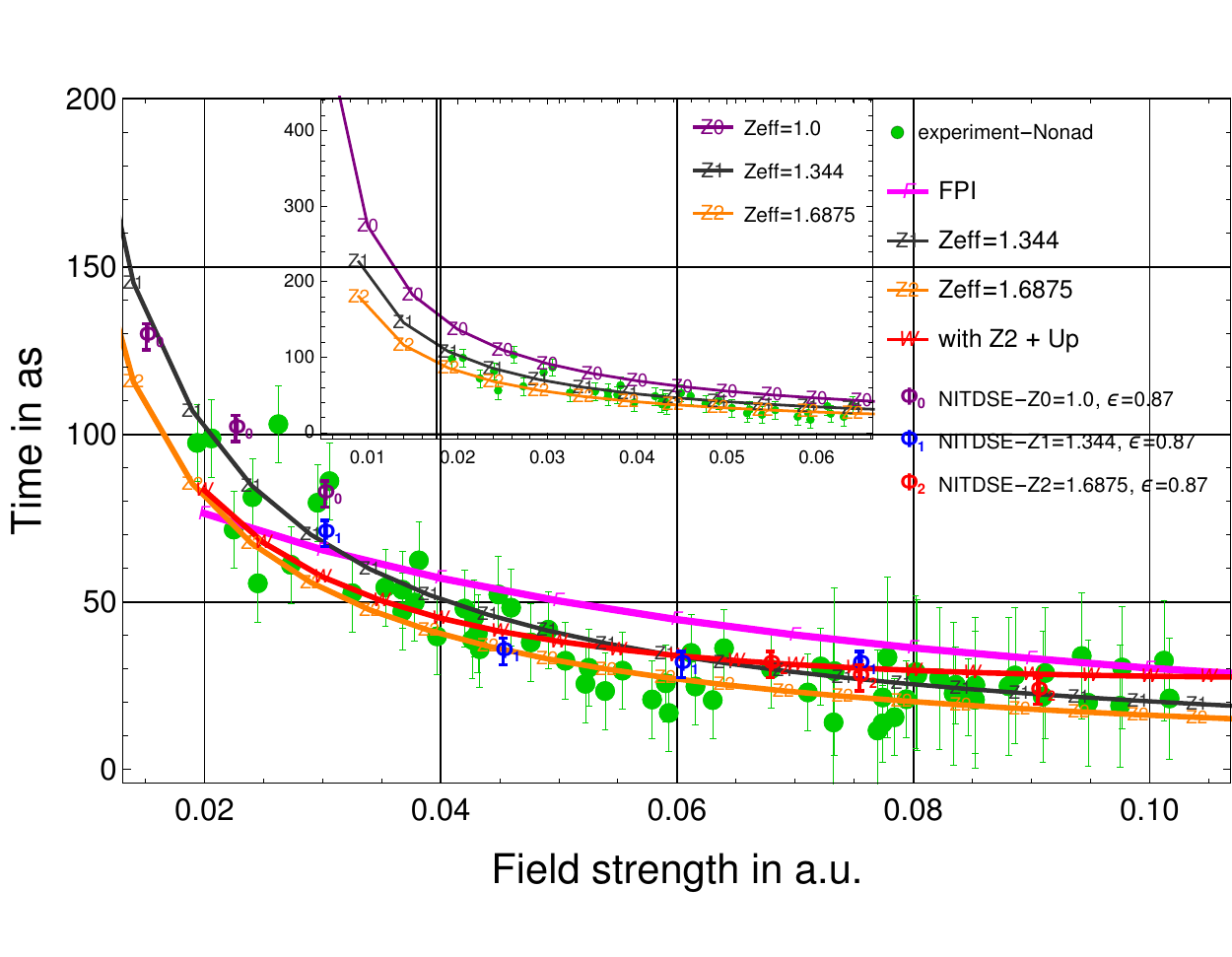}}
 \vspace{-0.60cm}
 \caption{\label{Hofdata3}\footnotesize (Color online)
 Graphic display of time delay versus field strength. 
 Plots as given in Fig. \ref{Hofdata2}  with  
 additional curves. W-curve (red) for $Z_{eff}=1.6875$ with $I_p\rightarrow I_p + 
 (\frac{F}{2\omega_0})^{2}\approx I_p+U_p$, where $I_p$
 is the   
 ionization potential of He atom and $U_p$ is the Ponderomotive potential,  
 see \cite{Delone:2000}. The FPI result (F-curve) 
 from \cite{Hofmann:2019}. 
 An enlarged range of the axes shows that our time delay $\tau_{dion}$ gives 
 perfectly the trend of the experimental data (see discussion in the 
 text). 
 In the inset the curves for three $Z_{eff}$  (from above)   
 $1.0, 1.344, 1.6875$. The NITDSE (see sec. \ref{sec:apdx} 
 and  \cite{IAIvanov:2014}: $\Phi_0$ (purple) $Z_{eff}=1.0$, $\Phi_1$ (Blue)  
 $Z_{eff}=1.344$  and $\Phi_2$ (Red) $Z_{eff}=1.6875$), see text.} 
 \end{figure}
 The intermediate regime is assumed to incorporate a coexistence of 
 tunneling and multiphoton ionization, and even the tunneling regime  
 is usually extended to cover $\gamma_K\sim 1$ \cite{Landsman:2014II} 
 and therefore is applied vaguely.   
  
 Fortunately, as we will see, we are able to account for the 
 nonadiabatic effects, where a two-step model is suggested, although 
 they are not strictly separated.
 In the first step, a scattering mechanism lowers the atomic potential 
 barrier by the amount $\varepsilon_F=I_p-\delta_z$, whereby the energy 
 is transferred to the electron (or electronic wave packet) by 
 a scattering mechanism.

 In the second step, traversing the barrier region (actually climbing 
 up the remaining barrier $\delta_z=I_p-\varepsilon_F$ or the effective 
 potential), is mainly compensated by a multiphoton absorption. 
 Where the intermediate regime, i.e. when traversing the barrier 
 region horizontally (horizontal channel) along with climbing the barrier 
 (climbing the vertical energy axis, vertical channel), will be discussed 
 in sec. \ref{sec:ir}.   
 We note that other nonadiabatic effects are small or negligible 
 as discussed by Hofmann et al. \cite{Hofmann:2019}. 
 Also, note that correlation or multielectron effects are negligible,  
 and the SAEA is valid for He-atom, as recently shown by Majety et al. 
 \cite{Majety:2017}.
 
 In our concept, a multiphoton absorption can be determined by the 
 barrier height $\delta_z$ (which depends on the field strength, see 
 Fig. \ref{Hofdata1}). 
 For now, the barrier height is largely overcome by a multiphoton 
 absorption after the scattering with the laser wave packet (first 
 step).    
 We note that in the SFA such a mechanism, i.e. when multiphoton 
 absorption accompanies the tunneling process, is usually incorporated 
 in the intermediate regime, which we discuss in sec.  \ref{sec:ir}.
 
 Our first, simplified approach follows from the physical reasoning of 
 $\tau_{T,i}$ in eq \ref{Tdi}.
 As already mentioned, $\tau_{T,i}$ is the time to reach the barrier 
 entrance $x_{e,-}$, in the adiabatic case, where the electron 
 encounters the barrier.   
 It can then climb the barrier by multiphoton absorption, it reaches 
 the top (sightly below the top) of the barrier and escapes the 
 effective potential at $x_m$, compare Fig. \ref{Hofdata1}.
 Consequently, the term due to the barrier itself is reduced by the 
 same amount of energy. 
 The number of absorbed (real) photons is then determined by the barrier 
 height which can approximated by $m_F=floor(\frac{\delta_z}
 {\omega_0})$, where $\omega_0$ is the central frequency of the 
 laser pulse; the function $floor(x)$ gives the  greatest integer 
 less than or equal to $x$. we use the $floor$ function instead of the  
 $ceil$, since the tiny remainder (=$\delta_z-m_F \omega_0<\omega_0$) of 
 the potential barrier can quietly overcome by (adiabatic) tunneling, 
 we cam back to this in sec. \ref{sec:ir}.  
 
 As the barrier energy reduces by the number of the absorbed photons, 
 that is by an amount $m_F\,\omega_0$, $\tau_{T,i}$ reduces as 
 follows,    
\begin{eqnarray}\label{dion1}\nonumber
 \tau_{T,i}(F)\rightarrow\tau_{dion}&=&\frac{1}{2}
 \frac{I_p-(\delta_z - m_F \omega_0)}{4 Z_{eff}F} \approx
 \frac{1}{2}\frac{I_p}{4 Z_{eff}F}\\
 &=& \frac{1}{2I_p}\frac{F_a}{F}=\frac{1}{2I_p}\zeta_{F}=\tau_a 
 \zeta_{F}
\end{eqnarray}
 Before we discuss the $\tau_{dion}$,  we first continue with our 
 analysis of the nonadiabatic tunneling process, which is certainly 
 much more complicated and richer with details than such a simple 
 reasoning \cite{Kullie:2018}.
 
 We can follow another point of view that is consistent 
 with the commonly applied SFA to calculate the T-time.
 As we shall see, the adiabatic and nonadiabatic calibrations differ  
 only by the $\tau_{delt}$ (see eq \ref{TdF}). 
 It turns out that the contribution $\tau_{delt}$ (the second term in 
 eq \ref{TdF}) is just due to the barrier itself, and is eliminated 
 by the nonadiabatic calibration.
 This is also in line with the velocity gauge, where a barrier does not 
 exist, which is important (compare Ni et al. \cite{Ni:20181}) and it is 
 in accordance with the restored equivalence of the two gauges in SFA 
 \cite{Ivanov:2005,Faisal:2007I,Faisal:2007II}.

 For the moment we assume that in the second step (after scattering 
 in the first step), that the multiphoton absorption amounts to 
 overcome the barrier and eliminate the second term in eq \ref{TdF}. 
 We note that in their work Klaiber et al. \cite{Klaiber:2016} 
 followed a similar reasoning, where the tunneling or the 
 field-ionization can be explained in a first step by a scattering 
 process, in which the atom is polarized by virtual multiphoton 
 absorption, or the wave function swells and the electron scatters 
 and gains some energy, which we can assume it corresponds to 
 our $\varepsilon_F=I_p-\delta_z$ (first step of our view).
 The second step by Klaiber el al. \cite{Klaiber:2016} is the tunneling  
 step from a virtual state $-I_p+n\,\omega_0$.
 The difference to the view of Klaiber et al. is that we have 
 $\, \nu_F=floor(\delta_z/\omega_0)(=m_F)$ instead of an (unspecified) 
 $n$ by Klaiber et al. and the multiphoton absorption to be similar. 
 eq \ref{TdF} becomes $\tau_{dion}$ because the second term 
 $\tau_{delt}$  in eq \ref{TdF} vanishes, as follows
\begin{eqnarray}\label{dion}\nonumber
 \tau_{dion}(F)&=&\frac{1}{2I_p}\frac{F_a}{F}+ 
 \bigg(\frac{1}{2I_p}\frac{F_a}{F}
 \frac{
 (\delta_z -{\nu_F} \omega_0)}{I_p}\approx0\bigg)\\
 &=& \frac{1}{2I_p}\frac{F_a}{F}=\frac{1}{2I_p}\zeta_{F}=\tau_a 
 \zeta_{F},
\end{eqnarray}
 which is the same result of eq \ref{dion1}, where $0\le {\nu_F}=
 floor\left(\frac{\delta_z}{\omega_0}\right)\le n_{I}=
 floor\left(\frac{I_p}{\omega_0}\right)$. 

 The presence of $Ip$ is inherent to energy conservation.
 It implies two steps, although not strictly separated, where an energy  
 transfer from the laser pulse to the electron by an amount   
 $\varepsilon_F$ by scattering (first step), or lowering the effective  
 potential below $I_p$ by the amount $\varepsilon_F$.  
 And a second step, in which multiphoton absorption occurs to overcome 
 the barrier (the remained effective potential) $\delta_z=I_p-
 \varepsilon_F$, but unlike eq \ref{dion1} virtual photons absorption 
 in eq \ref{dion} is, in principle, not excluded in addition to the 
 real photon absorption (furthermore in sec. \ref{sec:ir}), hence the 
 notation $\nu_F$ instead of $m_F$, although for the moment $\nu_F$ 
 represent a real photons number ($\nu_F\equiv m_F$).  
 
 Because $\tau_{T,i}$, $\tau_{T,d}$ leads to the same result 
 ($\tau_{dion}$), it becomes clear that the field-ionization time 
 delay is a real quantity and the nonadiabatic field ionization 
 follows directly from our adiabatic model eq \ref{Tdi}, where only 
 the energy conservation is required.
 If we consider both equations \ref{dion1}, \ref{dion} and realize 
 that according to the adiabatic case, they correspond to forwards, 
 backwards tunneling, the delay time is then the mean value 
 (symmetrization) and we obtain,     
\begin{eqnarray}\label{Tsym}
 \tau_{dion}(F)=\tau_{sym}&\equiv&\frac{1}{2}\left(\tau_{T,i}+ 
 \tau_{T,d}\right)=\\\nonumber
 &=&\frac{1}{2} \tau_{total}=\frac{1}{2I_p}\frac{F_a}{F}
\end{eqnarray}
 The factor $\frac{1}{2}$ is the symmetrization factor introduced in 
 our previous work \cite{Kullie:2020} to get the time delay from the 
 Aharonov-Bohm (ABTP) and Fujiwara-Kobe (FKTO) time operators, compare  
 sec. 3 in \cite{Kullie:2020}.
 Note symmetrization means an observable, i.e. physically a real 
 quantity, which presented by a Hermitian operator (such as ABTP or 
 FKTO, which are discussed in \cite{Kullie:2020}, see also \cite{Busch:2008} 
 and the references therein, and the time-of-arrival (TOA) distribution 
 of Simbillo et al. \cite{Sombillo:2018}, see discussion further below), 
 although it is still a controversial issue. 
 The result of eq \ref{dion}, \ref{Tsym} is fundamental and is 
 supported by the good agreement of $\tau_{dion}$ (or $\tau_{sym}$) 
 with the experimental result (despite the error bars) of Hofmann et al. 
 \cite{Hofmann:2019} as shown in Figs \ref{Hofdata2}, \ref{Hofdata3} 
 that we will discuss below, but let us discuss eqs 
 \ref{dion1}-\ref{Tsym} further first.  
 
 The contribution of the time delay $\tau_{delt}$ in eq \ref{Tsym} is 
 removed because $\delta_z$ cancels by the symmetrization, see details in 
 \cite{Kullie:2020}. 
 Under these considerations, it becomes unclear whether $\tau_{delt}$ is  
 eliminated by real or virtual multiphoton absorption. 
 However, eqs \ref{dion1}-\ref{Tsym} tell us that 
 the energy transfer can happen by either way.  
 Therefore, we suggest that  the energy gap of the neutral system (the 
 ionization potential) can be decomposed in the form 
 $I_p=v_F\, \omega_0 + n_F\, \omega_0$, where $v_F$, $n_F$ are 
 effective numbers of virtual and real photons, they are related to the 
 above-mentioned energy decomposition $\varepsilon_F, (I_p-\varepsilon_F)=\delta_z$, 
 respectively.
 Thus, we can assume that virtual photon absorption is equivalent to 
 a kinetic energy part (a scattering process or the polarization step 
 in the work of Klaiber et al. \cite{Klaiber:2016}), whereas climbing   
 the potential barrier is equivalent to real multiphoton absorption.  
 We can specify the energy proportions by $\varepsilon_F \approx 
 {v_F}\,\omega_0$ (scattering) and $(I_p-\varepsilon_F)=
 \delta_z \approx{n_F}\,\omega_0 =\nu_F\,\omega_0=m_F\,\omega_0$ 
 (multiphoton absorption), apart from the ponderomotive energy or 
 the Stark-shift, see below. 
 Still $\nu_F$ can include real and multiphoton parts, but it is 
 unnecessary to take this into account for the moment, i.e. we put 
 ${\nu_F}= m_F=floor(\delta_z/\omega_0)$.
 
 It is straightforward to examine the limits of this energy partition. 
 For $\lim_{F\to F_a}$ the barrier height disappears 
 $\delta_z\approx 0$ and $n_F=0$, $\varepsilon_F ={v_F}\,
 \omega_0 \approx I_p$, the kinetic energy transferred to the electron 
 from the pulse (suppressing the barrier) approximately equals $I_p$.   
 On the opposite side, for a small field strength ($\lim_{F\to 0}$) we 
 have $\delta \approx I_p, \varepsilon_F \approx 0$ and the $I_p\approx 
 n_{(F\to 0)}\,\omega_0=n_{I}\,\omega_0$ and the barrier is overcome by real 
 multiphoton absorption, which is equivalent to the well-know 
 multiphoton regime of the SFA for $\gamma_K\gg 1$.
 Therefore, we conclude that in the range $F\sim 0.02- 0.1$ of field 
 strengths used by the experiment of Hofmann et al. \cite{Hofmann:2019}, 
 both mechanisms are present, i.e. scattering and multiphoton absorption.
 We are aware that the process is complicated and different effects are 
 involved in the process. 
 Especially, a nonlinear Compton scattering mechanism is involved  
 \cite{Kullie:2018}, \cite{Eberly:1965}. 
 However, this is not critical since, firstly, their contribution is 
 small, and secondly, they correspond to a small kinetic energy 
 contribution, see the discussion further in sec. \ref{sec:dis}, 
 \,\ref{sec:cr}.
 Furthermore, a small (or tiny) tunneling contribution just below the 
 top barrier related to $\tau_{delt}$ is also possible as already 
 mentioned, which we will discuss in sec. \ref{sec:ir}.  
 
 We now come to the comparison with the experimental result. 
 The relation in eqs \ref{dion},\, \ref{Tsym} (and \ref{dion1}) shows 
 a very good agreement with the experimental data in the nonadiabatic  
 calibration of Hofmann et al. \cite{Hofmann:2019} as shown in Figs 
 \ref{Hofdata2}, \ref{Hofdata3}, where we plot $\tau_{dion}(F),\,  
 \tau_{sym}(F)$ for two values of $Z_{eff}$, together with the 
 experimental data of Hofmann et al.
 In the Fig. \ref{Hofdata2} the lowest curve (orange) for an effective 
 nuclear charge $Z_{eff}=1.6875$ of Clementi \cite{Clementi:1963}, and 
 an upper (gray) curve for $Z_{eff}=1.344=\sqrt{2 I_p}$ 
 \footnote{In previous works, O. Kullie used 
 $Z_{eff}=1.375 \approx\sqrt{2I_p}\approx 1.344$ from a model based on 
 an unpublished thesis of \\  
 him at the university of Kiel (Germany) for 
 the He-atom, where the full screening of the first electron is 
 accounted for. \\ 
 We still intend to expand the work for light rare gas atoms in the 
 framework of the SAEA and publish it whenever possible.}, where 
 $I_p$ the ionization potential of He-atom.     
 As seen, the difference between the two curves is smaller than the 
 error bars, thus, the value of $Z_{eff}$ is not crucial.
 In Fig. \ref{Hofdata3}, we also plotted $\tau_{dion}(F)$ for 
 $Z_{eff}=1.0$ (inset, purple) and a curve for $Z_{eff}=1.6875$ 
 (red, W-curve) by including the
 energy (continuum) shift given by $(\frac{F}{2\omega_0})^2$ 
 \cite{Delone:2000} (chap. 2, p.19), i.e. by replacing $I_p$ with 
 $I_p+ (\frac{F}{2\omega_0})^2$ ($\approx I_p+ U_p$, where $U_p$ the 
 Ponderomotive potential), which is negligible for $F<0.05$. 
 We also included in Fig. \ref{Hofdata3} the Feynman path integral 
 (FPI), (magenta, F-curve), from the same work of Hofmann et al.
 \cite{Hofmann:2019}.
 
 Furthermore, we  show  in the figures our calculated result of the 
 NITDSE, see sec. \ref{sec:apdx} and \cite{IAIvanov:2014},  for 
 $Z_{eff}=1.6875, 1.344$ in Fig. \ref{Hofdata2} and also for 
 $Z_{eff}=1.0$ in Fig.  \ref{Hofdata3}.  
 As seen in the figures, NITDSE is in an excellent agreement with 
 our $\tau_{dion}$ result and confirms our model.  
 
 Looking back to the NITDSE result of Ref. \cite{IAIvanov:2014}, one 
 finds  that the NITDSE was compared to experimental data of Boge et al. 
 \cite{Boge:2013} using a nonadiabatic calibration, see further below  
 sec. \ref{sec:cr}.
 The data of Boge et al. \cite{Boge:2013} (below Fig. \ref{Hofdata7}) 
 differs slightly (a bit higher) from the data of Hofmann et a.l 
 \cite{Hofmann:2019}, so that the NITDSE result of Ivanov et al. 
 \cite{IAIvanov:2014} was not close to the experimental data of Boge et  
 al. as it is the case in Fig. \ref{Hofdata2}, \ref{Hofdata3}, see also 
 the discussion in \cite{Kullie:2018}.   
 Finally, the good agreement between these results, our result, the 
 NITDSE, and the recent experimental data of Hofmann et al., shows that 
 our point of view offers a reasonable explanation of the issue. 
 
 Most likely $Z_{eff}=1.0$ fits better for small field strengths, 
 although no experimental result is available, compare Fig. 
 \ref{Hofdata3} (inset, upper curve). 
 The fact that $Z_{eff}=1.0, 1.344$ becomes closer to the curves of 
 $Z_{eff}=1.6875$ towards $F_a$ region is because $Z_{eff}$ has a lower 
 weight in the product $(Z_{eff} F)$ for larger $F$ values, but 
 one can easily see that the slope of $Z_{eff}=1.0$ curve does not 
 match to the experimental values in the region $F=0.2-0.5$.        
 Also note that despite the importance of the FPI result 
 \cite{Landsman:2014II} (see \cite{Landsman:2015}, \cite{Kullie:2015}), 
 it does not fit well with the experimental data, in particular the trend  
 is not satisfactory.
 In contrast to the flat behavior for larger field strengths, for 
 smaller field strengths $F\le 0.02$ the T-time becomes very steep with 
 a large slope, see Fig. \ref{Hofdata3}.
 One might even think that the (ionization) time delay $\tau_{dion} \, 
 (\tau_{sym})$ of eq \ref{dion} (eq \ref{Tsym}) is valid for small field 
 strengths.
 
 The good agreement of our result with the experimental data 
 as shown in figs \ref{Hofdata2}, \ref{Hofdata3} indicates that the 
 main behavior of the time delay is determined by the $\sim\frac{1}{F}$ 
 dependence, which is similar to the classical behavior or the Keldysh 
 time eq \ref{gamK} \cite{Kullie:2016}.  
 One also notices that both scale similarly with the ionization 
 potential, Keldysh time $\tau_K\sim \sqrt{2I_p}$ and our time delay 
 $\tau_{dion}\sim \frac{I_p}{Z_{eff}}\sim \sqrt{2I_p}$. 
 Apparently, $\tau_K$ is a classical quantity, while our 
 time delay is the corresponding quantum mechanical quantity. 
\subsubsection*{Note on one-Dimensional model}
Finally, we note that the agreement of our one-dimensional (1D) model 
with the three dimensional (3D) NITDSE is not surprising.
The 1D model (along with the 3D model) is widely used in the attoclock 
and attosecond science  \cite{Yakaboylu:2013,Ramos:2020,Klaiber:2020,
Kim:2021,Douguet:2018,Xu:2020,Yusofsani:2020,Canario:2022}. 
It is justified by the well known fact that (tunnel-) ionization in 
strong field regime occurs primarily along the direction of the 
electric-field at maximum, see \cite{Yakaboylu:2013,Canario:2022}.  
Then, the tunneling probability increases with decreasing width of the 
barrier ($\sim \frac{1}{F}$ dependence, $d_B=\delta_z/F$ see Fig. 
\ref{Hofdata1}); and the most probable tunneling path is concentrated 
along the electric-field direction at maximum.
The longitudinal contribution of the Coulomb potential into  
the dynamics represents the leading-order, while the transversal effect 
of the Coulomb potential is a higher-order correction 
\cite{Yakaboylu:2013,Canario:2022}.

In fact, the attoclock scheme \cite{Eckle:2008,Eckle:2008s,Trabert:2021}  
enables one to indicate the departure of the 1D model from the 
3D-model, where the 1D-axis is along the barrier width.
For a circular polarized laser pulse, 
the (tunnel-) ionization occurs primarily along the direction 
of the electric-field at maximum every half cycle, giving rise to 
a twofold symmetry of the electron momentum distribution in the 
polarization plane.   
The deviation of the orientation of the twofold symmetric distribution  
is referred to as angular offset \cite{Trabert:2021}. 
Thus, for a close to circular polarized laser pulse, the deviation 
from the twofold symmetric distribution depicts the deviation from the 
1D-model (tunnel-) ionization along the barrier width. 
Therefore, the asymmetry shown in Fig \ref{tdse}, which characterizes 
the orientation of the electric-field in the period of (tunnel-) 
ionization time (the angular offset), implies the small deviation of the 
1D-model (along the barrier width) from the 3D-model.

With this and similar to the note in \cite{Yakaboylu:2013}, we 
give an estimate of the deviation of the 1D-model from 3D-model.  
As seen in Fig \ref{tdse} (sec. \ref{sec:apdx}),   
$\frac{p_{x}}{p_{y}}\sim (0.25/1.5) \sim 17\%$  
(the propagation direction is along the z-axis). 
The minimum of the barrier width is reached at the atomic field 
strength $F_a$.  
In our case, we have $F_a \sim 0.12\, au$, which is the values 
used in the figure Fig \ref{tdse}. 
Accordingly, the deviation perpendicular ($xy-$ or ellipticity-plane) 
to the 1D-along-barrier ($z$-axis) is expected to be of a second 
order of the barrier width at $F \sim 0.12\, au$, 
and we could assume a similar percentage for lower field strengths. 
Now, keeping in mind that the distance along the path (curve) covered  
by the (tunnel-) ionized electron can be approximated by the 
simple arithmetic relation
\[d_{path} \approx \sqrt{d_{\parallel}^{2} + d_{\perp}^{2}}
=\sqrt{d_{\parallel}^{2} + 
((0.25/1.5)* d_{\parallel})^{2}} \sim 1.02\, d_{\parallel},  \]
where $d_{\parallel} (d_{\perp})$ is the 1D-distance along 
(perpendicular to) the barrier width, see Fig \ref{Hofdata1}.  
As we see, the effect is only about $\Delta d_{path}\sim 2\,\%$ for 
$F=0.12\, au$. 
Since the deflection angle is small, we can also use the arc 
$(s_{\phi} (F))$ (as calculated by NITDSE) and get for 
$Z_{eff}=1.6875$, $\Delta d_{path}(s_{\phi})\le 5\%\, d_{\parallel}$ 
(or $\lesssim 15\%\, d_{\parallel}$ for $Z_{eff}=1.0$) in the 
range of the applied field strengths.
Thus, we expect that the effect on the time delay is of the same 
order $\sim 5\% - 15\%$ for $Z_{eff}=1.6875-1.0$, respectively.  
It is a second-order correction, much like the claim of 
\cite{Yakaboylu:2013} mentioned above.  
The effect on the time duration for $Z_{eff}=1.6875$ is 
$\sim 5 \% \, \cdot \tau_d \approx 1 - 5\, as$ 
(or $\lesssim 3.5-20\, as$ for $Z_{eff}=1.0$) in the range 
$F=0.1 - 0.02$, respectively.  
It is about or less than the error bars given by the experiment 
and the error bars given by the NITDSE calculation of one degree 
(offset angle) $\sim 4\, as$ in the Figs \ref{Hofdata2}, 
\ref{Hofdata3}. 
It is worth noting that in strong field regime, the correlation 
do not have a significant effect on the offset angle 
\cite{Majety:2017}.  
To conclude, the 1D-model is justified in strong field regime 
and only for small field strengths less (or much less) than 
$0.01\, au$, the discrepancy between 3D- and 1D-model of the 
streaking angle is significant.    
\section{Discussion}\label{sec:dis}
 Looking to $\tau_{dion}=\tau_{sym}=\frac{1}{2}\frac{I_p}{4Z_{eff}F}$ 
 we see that $I_p$ determines the ionization time, 
 although the barrier height is $\delta_z$, compare Fig. \ref{Hofdata1}. 
 When the laser pulse (wave packet) scatters on the atom and 
 its field $F$ bends the atomic potential curve, the gain of the energy  
 $\varepsilon_F$ corresponds to the strength of the bending or lowering 
 the barrier below the continuum (first step), compare Fig. 
 \ref{Hofdata1}.
 Since, in such a case, it is usual to use the concept of virtual photon, 
 we write $\varepsilon_F$ in the form $\varepsilon_F\approx v_F\,
 \omega_0$, where $v_F$ denotes the number of the virtual photons. 
 
 Thus, due to the conservation of the energy, we can write $I_p =
 \varepsilon_F + \delta_z,\,\delta_z\approx n_F\omega_0
  (=\nu_F\omega_0=m_F\omega_0)$ (apart from a tiny remainder as 
  already mentioned and the Ponderomotive energy or the continuum 
  Stark-shift \cite{Delone:2000} p. 19, compare Fig. \ref{Hofdata3} 
  W-curve), where for the sake of simplicity we use the notation $n_F$. 
 $n_{F}$ ($0\le n_{F}=floor({\delta_z}/{\omega_0})\le n_{I}$) is the 
 (minimum) number of photons required to (climb) overcome the barrier 
 height (second step) at a field strength $F$, which is the barrier or 
 energy gap $\delta_z$ for the interacting electron \cite{Kullie:2016}, 
 compare Fig. \ref{Hofdata1}. 
 Therefore, the presence of $I_p$ in $\tau_{dion}$ indicates that both 
 steps take (real) time to happen, as it should be.  
 
 It is worthwhile to mention that our real time (delay) picture is consistent 
 with the approach discussed by Klaiber et al. 
 \cite{Klaiber:2015,Klaiber:2016} (furthermore in sec. \ref{sec:cr}), 
 whereas in the imaginary T-time picture or instantaneous tunneling, 
 see e.g. \cite{Sainadh:2019,Sainadh:2020}, the offset angle measured by 
 the experiment is attributed to the tail of the potential.  
 The first term $\varepsilon_F$ is implicitly considered, as the first 
 step (e.g. by Sainadh et al. \cite{Sainadh:2019}) and claimed to be 
 the collapse of the wave function in the orthodox interpretations of 
 the QM and to be in zeptosecond range \cite{Sainadh:2019}.  
 A similar conclusion is claimed by Ni et al. \cite{Ni:20181,Ni:20182}
 using classical back propagation. 
 One finds that the aforementioned imaginary T-time picture approach 
 (instantaneous tunneling) agrees with the adiabatic field calibration, 
 which can be compared with our adiabatic T-time picture $\tau_{_{T,d}}$ 
 (eq \ref{Tdi}), which agrees well with the experimental result for 
 H-atom \cite{Sainadh:2019} and the accompanied NITDSE result as  
 discussed in \cite{Kullie:20181}, apart from a factor $1/2$, which is 
 discussed in \cite{Kullie:2020}.
 
 In the perturbation regime where $F$ is small,  we have $\delta_z 
 \approx I_p$, $n_F=n_{I}$, whereas for a strong field one easily finds 
 that the number of absorbed photons is decreased by a factor that 
 depends on the field strength $0\le n_F \approx n_{I} \sqrt{1-F/F_a}$.  
 In other words, $n_F$ is the threshold number of photon required to 
 satisfy the energy conservation  with $I_p=\varepsilon_F+ \delta_z$ 
 $\left(\varepsilon_F\approx (n_{I}-n_F)\, \omega_0\right)$, which, i.e. 
 $n_F$, is not taken into account in the adiabatic tunnel-ionization (relative 
 to $I_p$  of the perturbation regime.)
 Then, $\varepsilon_F = I_p - \delta_z\equiv\Delta E$ can be used as an 
 energy uncertainty and by the virtue of the uncertainty principle, we 
 have $\tau_{T,d}=1/(2\Delta E)=1/(2(I_p-\delta_z))$, as given in eq 
 \ref{Tdi}, see \cite{Kullie:2016,Kullie:2015,Kullie:20181,Sainadh:2019} 
 for details. 
 For $F\rightarrow F_a$ we have $\varepsilon_F(\equiv\Delta E) 
 \rightarrow I_p, \, n_F=0$, the barrier $\delta_z(F=F_a)=0$ 
 vanishes and the delay time reaches its quantum limit $\tau_{T,d}=
 \tau_{dion}=1/(2 I_p)$, as already discussed. 
 At the opposite limit $F \rightarrow 0, n_F=n_{I}$, $v_F \rightarrow 0,\, 
 \varepsilon_F \rightarrow 0,\, \delta_z(F=0)=I_p$.
 In this case we have $\tau_{T,d} \rightarrow \infty$, $\tau_{dion} 
 \rightarrow \infty$ and the electron stays undisturbed in its ground 
 state. Further below in sec. \ref{sec:ir}. 
 
 It is worthwhile to mention that our model is consistent with the 
 result of Sombillo et al. \cite{Sombillo:2018} for the time 
 distribution of an incident particle using the  TOA formalism. 
 In fact, Sombillo et al. concern in their work the TOA operator and 
 suggest an interpretation of TEUR, which is in line with our point 
 of view. 
 They found that as the width increases, the traversal peak time moves  
 to higher values like our time delays, whereas in the opposite 
 direction, it results in a traversal-time distribution with a peak 
 shifted towards lower values of time, that is in our model when 
 $\tau_{T,d}$ reaches the quantum limits $1/(2I_p)$ at $F=F_a$, which 
 explains the Hartman effect. 
  
 With our model of eqs \ref{Tdi}--\ref{Tsym} we found a correspondence 
 between the time delay of adiabatic and the nonadiabatic 
 field-ionization, which agree well with the experimental results in 
 both cases of the field calibrations, the nonadiabatic and adiabatic, 
 for He atom \cite{Hofmann:2019}  and \cite{Landsman:2014II} (and  
 \cite{Sainadh:2019} for H-atom), respectively. 
 
 At this point, looking to eq \ref{TdF} (second line), it seems that 
 in the nonadiabatic calibration the time delay is determined by 
 $\tau_{dion}$ in the sense that the nonadiabatic effects beyond 
 multiphoton absorption are small, whereas the adiabatic calibration 
 is identified by the presence of a second term, the barrier term 
 $\delta/I_p$, which give $\tau_{delt}$.
 Then, the enhancement factor $\zeta_F=F_a/F$  
 is present in both cases and manifests the time delay $\tau_{dion}=
 \tau_a\zeta_F$, which is the (ionization) time delay in the 
 nonadiabatic calibration, precisely assuming that other 
 nonadiabatic effects beyond the multiphoton absorption are 
 negligible. 
 It corresponds to the self-interference term introduced by Winful 
 (eq \ref{Winf}) in the UTTP. 
 
 Considering the experimental data of Hofmann et al. \cite{Hofmann:2019} 
 to present the correct calibration (apart from the error bars)
 to the tunneling issue (or to the field-ionization and field 
 calibration in strong field region), we think that our explanation 
 provides a clear and comprehensive picture for the field-ionization 
 (or tunnel-ionization) in the strong field and the attoclock. 
 It enables us to describe very well both experimentally constructed 
 results for He- and H-atom, the adiabatic in \cite{Kullie:2015}, 
 \cite{Kullie:20181} and the nonadiabatic in the present work for 
 He-atom.
 Interestingly, our result shows an excellent agreement with the 
 NITDSE result as seen in figs \ref{Hofdata2}, \ref{Hofdata3}. 
 The agreement with the NITDSE greatly supports our model.
 Note that in eqs \ref{TdF}-\ref{Tsym} the enhancement factors 
 $\zeta_F, \, \Lambda_F,\, \chi_{_F}$ are relative dimensionless 
 factors, and we can use the intensity instead of the field strength, 
 $F/F_a=\sqrt{{I_L}/{I_a}}$, where $I_a$ is the appearance intensity 
 \cite{Augst:1989} and $I_L$ the intensity of the laser pulse. 
 
 From the result in eqs \ref{dion}-\ref{Tsym} and figs 
 \ref{Hofdata2}-\ref{Hofdata3} and our discussion so far, we conclude 
 that the interaction with the strong field can also be understood as  
 a combined process of a scattering ($\varepsilon_F$) and 
 a multiphoton absorption ($n_F \omega_0$), as far as nonadiabatic 
 calibration is concerned, e.g. as done by Hofmann et al. 
 \cite{Hofmann:2019}, where other nonadiabatic effects are small or 
 negligible, as already mentioned, which implies that energy gain 
 beyond multiphoton absorption can be neglected.
 The question is what or where is the difference between the weak and  
 strong field interaction processes, by neglecting the (smaller) 
 nonadiabatic effects beyond the multiphoton absorption (e.g. pulse 
 duration or change of field strength during the period of 
 traversing the barrier region or intensity fluctuations). 
 In fact, these effects are noticeably below the error bars. 
 Obviously, a main effect is the scattering process and polarization 
 of the electronic wave packet \cite{Klaiber:2016}, 
 \cite{Kullie:2018}, or the shrinking of the (energy) gap $I_p$ down, 
 it becomes $\delta_z(F)<I_p$ up to $\delta_z\approx 0$ at $F=F_a$, 
 whereas for small field strength $\delta_z(F\rightarrow 0)\approx I_p$.
 
 The time delay decreases with increasing field strength $F$ (and vice  
 versa), in accordance with the uncertainty principle 
 \cite{Aharonov:2000,Kullie:2016,Sombillo:2018}, and is determined by 
 the enhancement factor $\zeta_{F}$ (or $\chi(F)$ for the adiabatic 
 case), which becomes unity at $F=F_a$, at which the ionization time 
 reaches its lower quantum limit $\tau_{dion}(F_a)=\tau_a=1/(2I_p)$.  
 Nevertheless, as far as nonadiabatic effects are concerned, 
 we can imagine that the above-mentioned two steps happen  
 simultaneously.  
 Similarly, many authors, e.g. Ivanov et al. \cite{Ivanov:2005} and 
 Klaiber et al. \cite{Klaiber:2016}, \cite{Klaiber:2015},
 maintain the point of view that the energy gain can be thought 
 as of an absorption of photons during the tunneling process.
 In \cite{Klaiber:2015} the authors claim that in the nonadiabatic 
 regime the energy gain (including a multiphoton absorption) occurs 
 during the course of the under-the-barrier motion, where they describe 
 the nonadiabatic energy gain semi-classically (with a classical action) 
 \cite{Klaiber:2015,Klaiber:2016} see below Fig. \ref{Hofdata7},   
 we come back to this point later in sec. \ref{sec:ir}, \ref{sec:cr}. 
 
 The error bars of the experiment are large (see Fig. \ref{Hofdata2}, 
 \ref{Hofdata3}), which makes it harder to verify that after photons 
 absorption, tunneling occurs slightly below the top of the barrier.  
 Indeed, one can better understand this point by eliminating the other   
 nonadiabatic effects, e.g. due to laser pulse duration (envelope), 
 rotating of the laser field during the period of field-ionization and   
 intensity fluctuations, which can be responsible for the spread of 
 experimental points, see sec.  \ref{sec:ir}. 
 Hofmann et al.\footnote{C. Hofmann, private communication} 
 noted that between recording one distribution to the next, the laser 
 parameters, setup, temperature in the lab, .... might change and have 
 a slight influence, in principle the data point should also have error 
 bars for their F-axis-position.  
 By the multiphoton absorption $\sim n_F$, a tunneling mechanism can 
 happen just below the threshold, or slightly below the top of the 
 barrier as already mentioned, where $\frac{\delta_z}{\omega_0}$ is 
 usually not an integer, and the absorption of $n_F$ photons lets 
 a fairly small energy gap $\frac{\delta E}{\omega_0}=
 (\frac{\delta_z}{\omega_0}-n_F)<1$, which permits a tunneling 
 mechanism, we will discuss this further in sec. \ref{sec:ir}. 
 The interaction process is more complicated and a complex 
 scattering mechanism and a nonlinear Compton scattering can be 
 involved, where energy and momentum are transferred to the tunneled 
 or ionized electron by the scattering process, see \cite{Kullie:2018}.
 They are related to the characteristic of the interaction of the 
 electron with the intense laser field \cite{Meyerhofer:1997} by  
 $\sim \left(\frac{F}{\omega_0} \right)^2$ and $\sim \alpha 
 \left(\frac{F}{\omega_0}\right)^2$, respectively. 
 $\alpha=1/c$ ($c$ the speed of light in vacuum) is the fine structure 
 constant, which is equal to the strength of the interaction of the 
 photon with the electron. 
 We are aware that our result in eqs \ref{dion1}-\ref{Tsym} and 
 \ref{Tdi}, \ref{TdF} should be understood  as a well-estimated 
 result for the time delay, which could serve as a step for an 
 extension towards more sophisticated quantum mechanical treatment.   
 
 It is worthwhile to mention that many authors use a different 
 definition for the atomic field strength, e.g. 
 $F_{a}^{K}=k^3=(2I_p)^{3/2}$ \cite{Perelomov:1966,Klaiber:2015}, 
 which is related to the Keldysh parameter.
 It leads to the Keldysh time as we can see by the substitution 
 $F_a \rightarrow F_{a}^{K}$ in eqs \ref{dion}, \ref{Tsym},    
 \[\frac{1}{2I_p}\frac{F_{a}^{K}}{F} =\frac{1}{2I_p}\frac{k^3}{F}
 =\frac{2I_p}{2I_p}\frac{\sqrt{2I_p}}{F}=\frac{\sqrt{2I_p}}{F}=\tau_K\]
 It is well known that Keldysh time is too large, a classical quantity  
 and does not describe tunneling or field-ionization time (delay), for 
 details see \cite{Kullie:2016}.  
 This, however, shows that our time delay $\tau_{dion}, \tau_{sym}$ 
 (eqs \ref{dion1}-\ref{Tsym}) is directly connected to SFA, where $F_a$  
 (thus $\delta_z$ \cite{Kullie:2015}) represents the correct parameter 
 to determine the time delay while the atomic field strength is given 
 by $F_a={I_{p}^{2}}/{4 Z_{eff}}$ 
 \cite{Augst:1989,Augst:1991,Kullie:2015} regardless of the Keldysh 
 parameter $\gamma_K$. 
 Considering the field strengths given by the experimental results, 
 the Keldysh parameter is in the range of $\gamma_K\approx0.76-2.2$ in 
 the adiabatic \cite{Landsman:2014II} and in the range 
 $\gamma_K\approx 0.8-4.3$ in the nonadiabatic case 
 \cite{Hofmann:2019}. 
 Hence, despite its importance for the SFA, the Keldysh parameter 
 loses its significance in this regime, commonly refrred to as 
 the intermediate regime, see sec. \ref{sec:ir}. 
 
 In summary, in the strong field regime the nonadiabatic field 
 calibration can be understood by a scattering process combined with 
 a second step, which is essentially a multiphoton absorption.
 The number of absorbed photons can be approximated by $n_F$, 
 $0\le n_F=floor(\frac{\delta_z}{\omega_0})\le n_{I}{\omega_0}$. 
 The scattering process can be understood in a semiclassical sense 
 that the (electric) field of laser pulse bends the (atomic) 
 potential barrier, which reduces the energy gap from $I_p$ to 
 $\delta_z$, apart from the (continuum) Stark-shift \cite{Delone:2000} 
 and neglecting the small contribution of other nonadiabatic effects, 
 see sec. \ref{sec:ir}. 
 This picture is well-supported by the good agreement of our 
 $\tau_{dion},\, \tau_{sym}$ (eqs \ref{dion1}-\ref{Tsym})with the  
 experimental results as shown in figs \ref{Hofdata2}, \ref{Hofdata3} 
 in the nonadiabatic case of Hofmann et al. \cite{Hofmann:2019}. 
 And in the adiabatic case (eqs  \ref{Tdi}, \ref{TdF}) of 
 Landsman et al. \cite{Landsman:2014II}, as previously shown in 
 \cite{Kullie:2015} for He-atom and in \cite{Kullie:20181} for 
 H-atom. Our result is strongly supported by the NITDSE.
 For small $F$ the gap becomes close to the ionization potential 
 $\lim_{F\rightarrow 0} \delta_z=I_p$ and  $n_F=n_{I}$. 
 In the perturbation regime, i.e. a low-intensity (a relatively small 
 field strength) and a low-frequency ($floor(I_p/\omega_0)\gg 1$),    
 a (non-resonant) ionization happens by multiphoton absorption of 
 $\sim n_{I}$ photons (usually $n_{I}+1$ is used, or 
 the $ceil$ instead of $floor$ function \cite{Delone:2000}).   
 We think that eqs  \ref{dion}, \ref{Tsym} can be also valid in this 
 case or serve as a good approximation, as we can see from Fig. 
 \ref{Hofdata3}. 
 Finally, our model in the nonadiabatic case is related to the 
 adiabatic case and although it follows a simplified approach, it is  
 important because it enables us to provide a detailed (but not 
 sophisticated) picture of the strong field interaction with the laser 
 pulse, in accordance with the Winful UTTP, which is important for the 
 tunneling theory in general.   
 Indeed, this is one of the reasons why it makes sense to study the 
 adiabatic and nonadiabatic field-ionization together, as there exist 
 two field calibrations for the same experiment and system (He-atom).
 \begin{figure}[t]
 \vspace{-1.00cm} 
 \resizebox{8cm}{!}{\includegraphics{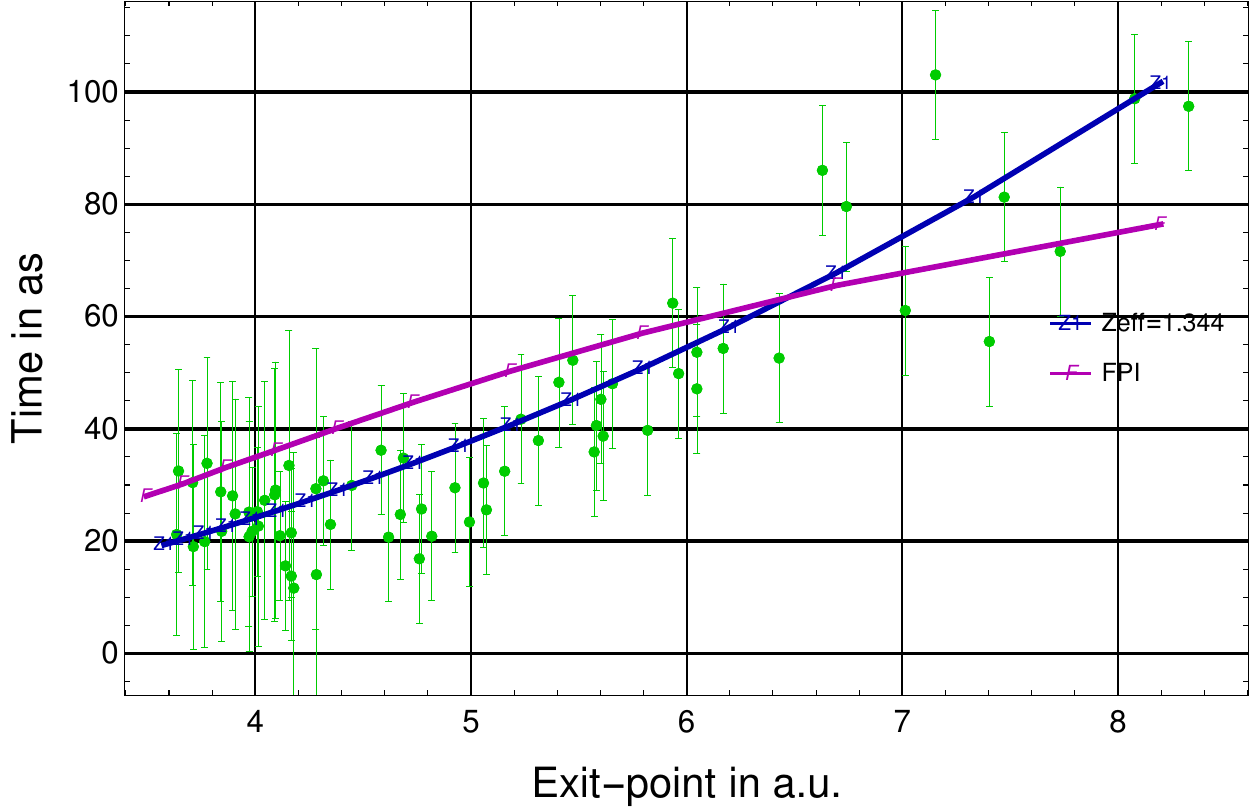}}
 \caption{\label{Hofdata4}\footnotesize (Color online)
 Graphic display data of Hofmann in the new calibration of the  field 
 strength, with our time delay $\tau_{dion}$ versus exit point 
 $x_m=\sqrt{Z_{eff}/F}$, with $Z_{eff}=1.344$  (a similar behavior is  
 found
 for  $Z_{eff}=1.6875$). F-curves denote FPI-curve.}
 \end{figure}
\section{The exit point}\label{sec:ep}
 It is common in the strong field and ultra-fast science to use the 
 so-called classical exit point $x_C=I_p/F$, see Fig. \ref{Hofdata1}, 
 to characterize the spatial location of the point at which the 
 tunneled or ionized electron escapes the potential barrier or the 
 effective potential, for details see \cite{Kullie:2016}. 
 Depending on the concept used to characterize the tunneling process, 
 it becomes free when it exits the ``exit'' point (`real' T-time 
 picture \cite{Kullie:2015}), or it becomes subject to the tail of the 
 potential (imaginary T-time picture \cite{Sainadh:2019}).
 A quick look at Fig. \ref{Hofdata1} shows immediately that 
 $x_C=I_p/F\equiv d_C$ is inaccurate and even wrong.
 For the adiabatic tunneling, it was shown in \cite{Kullie:20182} that 
 (in a semi-classical picture) the correct exit point is $x_{e,+}$ 
 (compare Fig. \ref{Hofdata1}) and the use of $x_C$ (or $d_C$) leads to 
 an erroneous conclusion.  
 In our nonadiabatic model (the present work), the exit point $x_{e,+}$ 
 is not suitable because multiphoton absorption is now involved, apart 
 from other nonadiabatic effects. 
 Thus, we expect a major effect on the exit point. 
 At first glance, this effect might result in an exit point equal to 
 $(1/2)x_C$ (eq \ref{xe} below), but as we will see, this may not 
 the correct approach and another approach is presented below.  
 
 Recalling what we did in eqs \ref{dion1}, \ref{dion} we find:
\begin{equation}\label{xe}
 x_{E} =\frac{I_p\pm\overbrace{(\delta_z- n_F 
 \omega_0)}^{\approx 0}}{2 F}
 =\frac{I_p}{2F}=\frac{1}{2} x_C, 
\end{equation}
 where the initial point is close to $x_i\sim 1 au$ (see fig \ref{Hofdata1}), 
 it is small and can be fairly neglected.
 We note that with eq \ref{xe} the barrier width vanishes 
 $d_B=x_{e,+}-x_{e,-}=0$, 
 whereas the traversed distance in this case is 
 $d_h=x_{exit}-x_{initial}\approx x_E$.   
 Nevertheless, because $x_E=x_C/2$ the overall picture is similar to 
 the case of the adiabatic calibration, see Fig. 2 of 
 \cite{Kullie:20182}. 
 This shows a linear dependence of the time delay versus the exit 
 point (the same $\sim \frac{1}{F}$ dependence of eqs \ref{dion}, 
 \ref{xe}), which is most likely unsuitable for such a process.
 
 The multiphoton absorption is usually depicted as a vertical 
 channel \cite{Ivanov:2005}, i.e. the electron climbs the effective 
 potential and moves towards its maximum. 
 Hence, we can characterize the exit point in this way.
 As seen in Fig. \ref{Hofdata1}, the maximum of the barrier height is 
 located at $x_m=\sqrt{{Z_{eff}}/{F}}$ and we expect that the 
 ionized electron, climbing the barrier, will moves towards $x_m$.
 In Fig. \ref{Hofdata4}, we plot the ionization time $\tau_{dion}$ 
 versus the exit point for the two $Z_{eff}=1.344,1.6875$.
 Unlike $x_E$, the exit point $x_m$, and hence, the curves in Fig. 
 \ref{Hofdata4} depend on $Z_{eff}$.
 The difference to the former case (linear dependence) is not 
 remarkable, although $x_m$ is noticeably smaller. 
 In addition, it is difficult to judge from Fig. \ref{Hofdata4} whether  
 $x_m$ actually determines the exit point. 
 Nevertheless, from the good agreement of $\tau_{dion}, \tau_{sym}$ 
 with the experimental data (as seen in Fig. \ref{Hofdata2}) for 
 $Z_{eff}$ values larger than $1.0$, i.e. $Z_{eff}=1.6875, 1.344$, we 
 think that the traversed distance by the ionized electron should not 
 be too large (not too far from the nucleus.)   
 Since $x_m$ is smaller than $x_E$ (eq \ref{xe}), the actual exit point 
 is most likely close to $x_m$, see further below sec. \ref{sec:ir}.
 This is unlike the adiabatic case, where no photon absorption is 
 involved in the tunneling process. 
 In the later case, the horizontal channel dominates the process of 
 tunneling (tunnel-ionization) \cite{Ivanov:2005}.             
 Finally, the so-called classical exit point $x_C$ is by no mean  
 a correct choice (compare Fig. \ref{Hofdata1}), see also 
 \cite{Kullie:20182} for the adiabatic case.
 Indeed, it is easy to see from the barrier width, 
 $d_B=\frac{\delta_z}{F}=\frac{I_p}{F}\sqrt{1-4Z_{eff} F/I_{p}^{2}}
 =x_C\sqrt{1-F/F_a}$  that $x_C$ is modified by 
 a factor that becomes approximately unity for small field strength,  
 $\lim_{F\to 0} \sqrt{1-F/F_a}\rightarrow 1$. 
 Hence, the so-called classical barrier width is justified only for 
 $F\ll F_a$ ($\gamma_K\gg  1$). 
 To conclude, it is difficult to determine the ``correct'' exit point 
 with our approach. 
 In the nonadiabatic case, one finds that the exit point of the 
 field-ionized electron is close to $x_m=\sqrt{Z_{eff}/F}$, but 
 $x_E=I_p/(2F)=(1/2)x_C$ is also  not excluded. 
 In any case, it is in a range between $x_m$ and $x_E$, see further 
 below, and not $x_C=I_p/F$ as usually done, e.g. \cite{Hofmann:2019}. 
\section{The intermediate regime}\label{sec:ir}
 In our model we have treated so far two experimentally given cases, 
 the nonadiabatic field calibration for He-atom in the present work 
 and the adiabatic field calibration case for He atom in 
 \cite{Kullie:2015} and for H-atom in \cite{Kullie:20181}. 
 In both cases, we found a good agreement with the experimental result. 
 The field calibration of Hofmann et al. \cite{Hofmann:2019} affects 
 a shift to a lower intensity. 
 It causes a shift of the time delay to a smaller value for the 
 same field strength.
 This confirms our tunneling model as seen in eq \ref{TdF},\ref{dion}, 
 since the second term vanishes $\tau_{delt}=0$, when both a scattering 
 and multiphoton absorption process are involved.  
 A feature of the experimental data (both adiabatic and nonadiabatic,   
 see below Fig. \ref{Hofdata6}) is the spread of the points.  
 This can be for a variety of reasons, such as pulse length or carrier 
 envelope phase. 
 However, as we will see below, this can also be caused by intensity 
 fluctuations that allow a tunneling contribution slightly below the 
 top of the barrier.
  
 The two cases of the field calibration can also be viewed as two 
 limits to the field-ionization process.
 This immediately raises the question about what is usually called 
 the intermediate regime, in which both a tunneling contribution and 
 a multiphoton absorption exist \cite{Ivanov:2005}.
  
 In deriving $\tau_{dion}$ ($\tau_{sym}$) we assumed that the number of 
 absorbed photon and the gain of the energy due to the scattering with 
 the laser wave packet, preserves the energy conservation $Ip \approx 
 \varepsilon_F + n_F \omega_0$ (approximately as 
 $n_F=floor(\delta_z/\omega_0)$), apart from the Stark-shift 
 \cite{Delone:2000} or the Ponderomotive energy. 
 Nevertheless, due to the complexity of the process the above-mentioned   
 two steps are not strictly separated, and the electron can also escape 
 by following a horizontal channel \cite{Ivanov:2005} while it climbs 
 up the energy axis (vertical channel), and ends up with an energy 
 $(\varepsilon_F+\varepsilon_{\omega_0}) \sim I_p$, regardless the 
 relative number of the (virtual and real) absorbed photons. 
 More specifically, we can write $I_p=(\varepsilon_F+
 \varepsilon_{\omega_0})+ \Delta\varepsilon$. 
 For $\Delta \varepsilon \approx 0$ we have $(\varepsilon_F+
 \varepsilon_{\omega_0})=\varepsilon_F+n_F \omega_0
 \approx(\varepsilon_F+\delta_z)\approx I_p$ corresponds to 
 $\tau_{dion}, \tau_{sym}$ of eqs \ref{dion}, \ref{Tsym}.
 Note, that the absorption of photons number larger than $n_F$  
 required by the energy conservation, is equivalent to an 
 above-threshold ionization (ATI) process, similar to the well known 
 non-resonant ionization  \cite{Helm:1994} in the perturbation regime. 
 In the following, we do not consider the ATI process, with the 
 possibility of addressing it in a future work.
 
 However, $(\varepsilon_F +\varepsilon_{\omega_0})+\Delta 
 \varepsilon>I_p$ (apart from $U_P$) means that the energy gain becomes 
 larger than the maximal barrier height $I_p$ and the electron escapes 
 with a velocity larger than zero.
 Therefore, in accordance with the SFA that the momentum peaks around 
 zero velocity, we assume that $\Delta \varepsilon$ is small.   
 In a first approach we approximate $\Delta\varepsilon$ and set 
 $\Delta \varepsilon\approx \Delta n\,\omega_0, \Delta 
 n=floor(\frac{\Delta \varepsilon}{\omega_0})$, where $\Delta n$ 
 is small compared to $n_F=floor(\frac{\delta_z}{\omega_0})$.
 eq \ref{dion}, \ref{Tsym} then becomes
\begin{eqnarray}\label{tion}\nonumber
\tau_{tion}(F)&=&\frac{1}{2}\frac{Ip+\Delta \varepsilon}{4 Z_{eff}F}
\\\nonumber
 &=& \frac{1}{2I_p}\left[\frac{F_a}{F} 
  \, \left(1+\frac{\Delta n\, \omega_0}{I_p}\right)\right]\\  
 &=& \tau_a \,\eta\left(F,\omega_0,\Delta n\right)
\end{eqnarray} 

 In eq \ref{tion} the gain of the energy happens while the electron is 
 non-adiabatically field-ionized by absorbing a number of photons  
 corresponds to $n_F +\Delta n$.
 We mention that a similar view is presented by Camus et al. 
 \cite{Camus:2017}.  
 
 Likewise, as discussed in sec. \ref{sec:tt} when we obtained eq 
 \ref{dion}, another point of view can be considered, in which
 the electron (or the electron wave packet) tunnels by absorbing an 
 effective number of photons $\tilde{n}_F < n_F$ so that $\tau_{delt}$ 
 does not vanish (compare eq \ref{dion}), while the first term 
 (self-interference term according to Winful UTTP) is preserved by the 
 virtue of the energy conservation. 
 In this case, we have $\varepsilon_F + \tilde{n}_F\,\omega_0 + 
 \Delta\epsilon\gtrsim I_p$, with an effective number of real photons 
 $\tilde{n}_F\lesssim n_F$. 
 $\Delta \epsilon$ is small, in accordance with the SFA, where 
 $\Delta \epsilon$ corresponds to a tunneling contribution 
 (horizontal channel).  
 With $\Delta \tau_{delt}=\frac{\Delta\epsilon}{4Z_{eff} F}$, and 
 $\Delta\epsilon=\Delta \nu\,\omega_0$ ($\Delta \nu \approx 
 floor(\frac{\Delta\epsilon}{\omega_0})$) and from eq \ref{TdF} or 
 eq \ref{dion}, we obtain 
\begin{eqnarray}\label{tion1}\nonumber
\tau_{tion}(F)&=& \tau_{dion}+\Delta\tau_{delt}\\\nonumber
 &=&\frac{1}{2I_p}\left[\frac{F_a}{F} 
    \, \left(1+\frac{\Delta \nu\, \omega_0}{I_p}\right)\right]\\  
 &=& \tau_a \,\eta\left(F,\omega_0,\Delta\nu\right)
\end{eqnarray} 
 And again $\Delta\epsilon=0,\Delta\nu=0$ corresponds to eq \ref{dion} 
 or \ref{Tsym}.   
 In eq \ref{tion1} the energy gain happens by absorbing $\tilde{n}_F$ 
 photons followed by a small tunneling contribution a little below the 
 top of the barrier.
 This is similar to the view of Klaiber et al. \cite{Klaiber:2015}, 
 more on this in  sec. \ref{sec:cr}.
 
 At first glance, we can imagine that an energy gain occurs during 
 the entire process, where horizontal and vertical channels coexist 
 \cite{Ivanov:2005}.  
 However, if one imagines that such a process takes place in a complex 
 mechanism, in which an energy gain occurs even after it tunnels/escapes 
 the exit point, we are led to the imaginary T-time picture discussed 
 by Sainadh et al. \cite{Sainadh:2019} and recently \cite{Sainadh:2020}.
 Note that for $\Delta \nu\rightarrow\nu_F$ ($\Delta \tau_{delt}
 \rightarrow \tau_{delt}$), eq \ref{tion1} becomes identical to eq 
 \ref{TdF} (or $\tau_{T,d}$ in eq \ref{Tdi}), which is the adiabatic 
 case. 
 We emphasize that our (real) T-time picture in the adiabatic case 
 agrees well with the imaginary T-time picture for H-atom and with the 
 NITDSE \cite{Kullie:20181,Sainadh:2020}. This undoubtedly confirms 
 our view. 

\begin{figure}[t]
\resizebox{8cm}{!}{\includegraphics{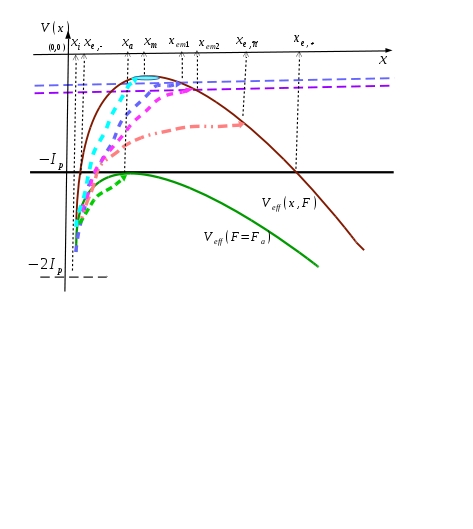}}
\vspace{-4.0cm}
\caption{\label{Hofdata5}\footnotesize(Color online)
 Illustration of the intermediate regime, eqs 
 \ref{tion1}-\ref{xetion1}.  The (orange) dashed dotted curve 
 illustrates the view of Klaiber et al. \cite{Klaiber:2015}, which is usually called 
 the nonadiabatic or intermediate tunneling 
 regime \cite{Ivanov:2005}, see text eqs  \ref{tion1}, \ref{xetion1}. 
 See  also Fig. \ref{Hofdata1}.}
\vspace{-0.0cm}
\end{figure}

\begin{figure}[t]
\resizebox{8.cm}{5.0cm}{\includegraphics{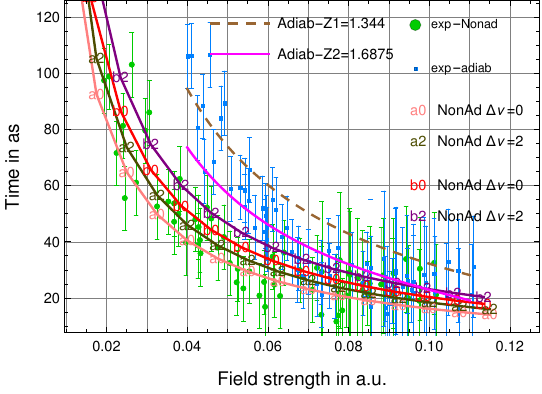}}
\caption{\label{Hofdata6}\footnotesize (Color online)
 The figure shows the time delay $\tau_{tion}$ given in eq \ref{tion1} 
 for $Z_{eff}=1.6875$ and $\Delta \nu=0,2$
 (lower two a-curves). 
 And $Z_{eff}=1.344$ and  $\Delta \nu=0,2$ (upper two b-curves).
 The experimental data (blue, 
 rectangles) with the adiabatic 
 \cite{Landsman:2014II} and (green, circles) with nonadiabatic 
 \cite{Hofmann:2019} field calibration.
 The two curves
 over the adiabatic experimental data are the T-time 
 $\tau_{_{T,d}}$ of eq \ref{Tdi} for $Z_{eff}=1.6875$  (below, 
 magenta),  $Z_{eff}=1.344$ (above, dashed light brown).}
\end{figure}

 As seen in eqs \ref{tion1} (eq \ref{tion}), the time delay increases 
 for $\Delta\nu>0$ (or $\Delta n>0$) and becomes larger than the 
 field-ionization time delay given by $\tau_{dion}, \tau_{sym}$ (or 
 the self-interference term $\tau_{si}$ according to eq \ref{Winf} in 
 the UTTP of Winful \cite{Winful:2003}).  
 With eq \ref{tion1} the situation now becomes similar to so-called
 intermediate tunneling regime, where the vertical and horizontal 
 channels co-exist \cite{Ivanov:2005}.
 Eq \ref{tion} becomes identical with eq \ref{tion1} by the 
 replacement $\Delta  \nu \rightarrow \Delta n$, where 
 eq \ref{tion1} is suitable (with $\Delta  \nu$ virtual photons number) 
 to describe a tunneling mechanism \cite{Klaiber:2016} and, as we will 
 see, to explain that the time delay increases by moving from the 
 nonadiabatic towards an adiabatic field calibration or adiabatic 
 tunneling, which is significant for the tunneling theory. 
 And, as we will see, it explains the spread of the experimental 
 points, which can be caused by intensity fluctuations of the laser 
 pulse.  
 Therefore, we restrict our discussion to eq \ref{tion1}.
 For tiny $\Delta\epsilon$ the tunneling through the horizontal channel 
 happens just under the top of the barrier, as depicted in Fig. 
 \ref{Hofdata5}, where the tunneling probability is notably high. 

 We summarize the time delay as the following
\begin{eqnarray}\nonumber
 \tau =\left\{
 \begin{array}{ll}
 {\tau_{dion}=\tau_a \, \zeta(F)}&
 \mbox{\rm{eqs  \ref{dion}, \ref{Tsym}, nonadiabatic}}\\  
 {\tau_{T,d} =\tau_a \, \chi(F)}& 
 {\rm{eqs  \,\, \ref{Tdi}, \ref{TdF},\,\, adiabatic}} \\    
 {\tau_{tion}=\tau_a \, \eta (F,\omega_0,\Delta\nu)}&
 {\rm{eq\,\,\, \ref{tion1},\,\,\, ``intermediate"}}
\end{array}\right.
\end{eqnarray}
 The summary is detailed in table \ref{timt}.
 With this, we can treat the intermediate regime, which  is generally  
 considered to be multiphoton absorption during tunneling.    
 It was first described by Ivanov et al. \cite{Ivanov:2005} and the 
 conclusion was that the two channels (horizontal and vertical) do not 
 exclude each other. 
 According to Ivanov they co-exist in a `gray' area $\gamma_K 
 \approx 1$, called `nonadiabatic tunneling'. 
 Similarly, Klaiber et  al. in \cite{Klaiber:2016} presented a view 
 that can be compared with our view in the following.  
 In our nonadiabatic picture, the horizontal channel (tunneling) is 
 a little below the top of the barrier.
 In Fig. \ref{Hofdata5}, we show an illustrative picture of two 
 intermediate cases, in which the multiphoton absorption is followed by  
 a tunneling from two intermediate virtual states below the top of 
 the barrier (the two horizontal dashed lines in Fig. \ref{Hofdata5}).  
 We illustrate this with the two dashed curves blue and purple, from 
 above the second and third (dashed) curves under the barrier, with 
 the exit points $x_{em1}, x_{em2}$ (see below), respectively.
 Whereas the highest dashed curve (light blue, from above the first 
 dashed one), illustrates the case of a negligible tunneling 
 contribution (where $\delta_z-n_F\omega_0<\omega_0$), with the exit 
 point $\approx x_{m}$. 
 A lower dashed-dotted curve (orange, from above the fourth one curve)   
 corresponds to the view of Klaiber et al. \cite{Klaiber:2015}, 
 according to which the tunneling happens significantly below the top 
 of the barrier. 
 We will come back later to this in sec. \ref{sec:cr}. 
 
 In addition, the limit case for $F=F_a$ is shown in Fig \ref{Hofdata5},  
 green dashed curve (the lowest curve with the exit point 
 $x_m(F_a)=x_a$).   
 In this case, the atom is highly polarized that the barrier disappears 
 and the BSI sets up. 
 The time to reach the entrance point $x_a$ (which coincides with the 
 exit point) is the quantum limit $\tau_a$, see eqs \ref{dion}, 
 \ref{Tsym}.  
 This picture agrees well with the scattering and the collisional 
 rearrangement process in the ion-atomic collision 
 \cite{Klaiber:2016,Briggs:1990}. 
\begin{table}
\footnotesize
 \begin{tabular}{|l|l|l|}\hline
 Regime       & Time delay expression & Enhancement expression \\\hline
 Nonadiabatic & $\tau_{dion} = \tau_{sym}=\frac{1}{2I_p}\frac{F_a}{F}$
              & $\tau_a  \zeta(F)$\\
 Adiabatic    & $\tau_{T,d}=\frac{1}{2I_p}\frac{F_a}{F} 
     \left(1+\frac{\delta_z}{I_p}\right)$ & $\tau_a \chi(F)$ \\
              &$\quad\quad\,=\tau_{dion}+\tau_{delt}$ & $= \tau_a \left(\zeta(F)+ \xi(F)\right)$\\
 Intermediate & $\tau_{tion} = \frac{1}{2I_p}\frac{F_a}{F} 
                \left(1+\frac{\Delta \nu\,\omega_0}{I_p}\right)$& $\tau_a\,\eta (F,\omega_0,\Delta\nu)$\\
              & $\quad\quad\,\,\,=\tau_{dion}+\Delta\tau_{delt}$&\\\hline
 \end{tabular}\\\vspace{0.2cm}
 \caption{\label{timt} \footnotesize The table summarizing the regimes and
 their associated time delay expressions.}
\end{table}
 As already mentioned, a nonlinear Compton type scattering with laser 
 pulse is involved, as experimentally investigated by Meyerhofer et al. 
 \cite{Meyerhofer:1997} and earlier in a theoretical work of Eberly et 
 al. \cite{Eberly:1965}.
 It is a collective scattering with the laser wave packet at high 
 photon density or strong field, where the electron recoils or 
 the electronic density is strongly polarized due to the strong 
 electric field of the laser \cite{Kullie:2018}, similar to the 
 ion-atom collision, as also discussed by Klaiber et al. 
 \cite{Klaiber:2016}.  
 Note, the effect caused by an electric field or a charge density is   
 the same. 
 According to Einstein, Wheeler and Feynman, electric charge and field 
 are the same and not independent entities \cite{Mead:2000,Mead:20131}.
 
 In Fig. \ref{Hofdata6}, we plot $\tau_{tion}$ of eq \ref{tion1} for 
 $Z_{eff}=1.6875$ (lower two curves) and $Z_{eff}=1.344$ (higher two 
 curves), where, for a better visibility, the curves correspond to 
 $\Delta\nu=0,2$ are plotted.
 In Fig. \ref{Hofdata6}, we included the result of the adiabatic case 
 \cite{Kullie:2015}, see eq \ref{Tdi}, with the experimental data of 
 Landsman et al. \cite{Landsman:2014II}.
 This result may explain one of the reasons (intensity fluctuation), 
 which causes the spread of the experimental points.
 It corresponds to the absorption of a slightly smaller number of 
 photons than required by $\delta_z$ ($\tilde{n}_F\lesssim n_F$) as 
 shown in Figs \ref{Hofdata5}, \ref{Hofdata6} and eq \ref{tion1}, 
 although it is difficult to ensure such a conclusion since the error 
 bars are larger than the separation between successive curves (or 
 even between $\Delta\nu=0, \Delta\nu=2$).
 In addition, fig \ref{Hofdata6} suggests that the $\tau_{dion}$ curves 
 (the nonadiabatic case) move towards $\tau_{T,d}$ curves, when 
 $\Delta\nu\, \omega_0$ become larger, see eq \ref{tion1}, up to 
 $\Delta\nu\,\omega_0=\delta_z$ (the adiabatic case). 
 
 We come to the exit point in the case of eq \ref{tion1}. 
 As discussed in sec. \ref{sec:ep}, it is in the range between $x_m$ 
 and $x_E$ in the nonadiabatic case (no tunneling contribution). 
 Whereas in the adiabatic case (adiabatic tunneling), it is estimated 
 by $x_{e,+}$, compare Fig. \ref{Hofdata5}. 
 In the case of eq \ref{tion1} (intermediate case), an approximate 
 value can be obtained by the same procedure applied to obtain eqs  
 \ref{tion}, \ref{tion1}.
 The exit point shifts from $x_{m}$ towards $x_{e,m1}, .., 
 x_{e,m_k}, ..$ for $\Delta\nu = 1,.., k, ..$, and reaches $x_{e,+}$ 
 for $\Delta\nu\,\omega_0=\nu_F\,\omega_0\approx \delta_z$ (i.e. 
 $\Delta\nu=\nu_F$ in eq \ref{tion1}). 
 The barrier width changes in the same way. 
 From the intersection points of the horizontal dashed lines (virtual 
 states) with the effective potential curve, in Fig. \ref{Hofdata5}, 
 we find  
\begin{eqnarray}\label{xetion}
  d_{B}^{\nu}&\approx& \frac{\Delta\epsilon}{F}
 \approx\frac{\Delta\nu\omega_0}{F}, 
\end{eqnarray}
 which becomes $d_B=\delta_z/F$ for $\Delta\nu=\nu_F$. 
 We obtain the exit point (compare Fig. \ref{Hofdata5}) approximately by,   
 \begin{equation}\label{xetion1}
  x_{e,m_k}\approx x_m +\frac{d_{B}^{\nu}}{2}=
  \sqrt{\frac{Z_{eff}}{F}} + \frac{\Delta\epsilon}{2F}
 \end{equation}
 Note, for $\Delta\nu =0,\, \Delta\epsilon=0$, we have $d_B^{\nu}=0$ 
 and $x_{em0}\equiv x_m$.  
 In this case, the field-ionization happens along the vertical channel 
 and the tunneling contribution is negligible, as already mentioned 
 (light blue curve, the first dashed one below the effective potential 
 curve in Fig. \ref{Hofdata5}.)
 Therefore, as we have seen in eq \ref{tion1} the second term in eq 
 \ref{xetion1} indicates the tunneling contribution. 
 The interesting case is tunneling near the top of the barrier, 
 that is when $\Delta\epsilon$ is small enough ($\Delta\nu\sim 0,1,2$), 
 where the tunneling probability is quite high, compare Fig. 
 \ref{Hofdata5}. 
 In our view, the spread of the experimental points can be traced back 
 to this issue.
 Furthermore, we see from eq \ref{xetion1} that the difference in the 
 number of absorbed photons change the exit point from $x_{m}$ to 
 $x_{em_1}, x_{em_2}, \cdots$ toward $x_{e,+}$ (reaches $x_E$ in 
 between), which increases the time delay from the nonadiabatic case 
 $\tau_{sym}, \tau_{dion}$ for $\Delta\nu=0$ (eq \ref{dion}, 
 \ref{Tsym}) towards the adiabatic case $\tau_{T,d}$ for $\Delta\nu=
 \nu_F$ (eq \ref{TdF} or \ref{Tdi}). 
 
Eq \ref{tion1} can be rewritten in the form 
\begin{equation}\label{tion2}
 \tau_{tion}(F)=\frac{1}{2I_p}\frac{F_a}{F}\left(1 +
 \frac{\Delta\nu\omega_0}{I_p}\right)
 = \tau_a \frac{F_a}{F}\left(1+\frac{\Delta\nu}{n_{I}} \right)
\end{equation} 
 Eq \ref{tion2} is important, since it is valid for the intermediate 
 tunneling, but is independent of the laser frequency $\omega_0$. 
 For $\Delta\nu=0$, we have eqs \ref{dion}, \ref{Tsym} (or the 
 self-interference term in the Winful tunneling model, eq \ref{Winf}).  
 It sets a limit from below to the time delay for the field-ionization, 
 the nonadiabatic case, where only a negligible tunneling contribution 
 exists (slightly below the barrier, the first dashed curve one from 
 above, light blue, in Fig. \ref{Hofdata5}.) 
 The second term in eq \ref{tion2} appears when $\Delta\nu>0$, which 
 is smaller than the first one and indicates a tunneling part, which 
 increases the delay time up to the adiabatic case at $\Delta\nu=\nu_F$,
 or precisely at the (maximum) barrier height $\Delta \epsilon =\delta_z$.
 
 In summary, after a first step or the scattering with the laser pulse, 
 the multiphoton absorption (vertical channel) can be followed by (or 
 co-exist with) a tunneling (horizontal channel, $\Delta\nu\ne 0 $ in 
 eq \ref{tion1}, \ref{tion2}), slightly below the top of the barrier, 
 where the tunneling probability is notably high, compare Figs. 
 \ref{Hofdata5}, \ref{Hofdata6}.  
 The amount of this contribution is smaller than the error bars in the 
 data of Hofmann et al. \cite{Hofmann:2019} (see Fig. \ref{Hofdata6}). 
 Therefore, in our view, a refinement on the experimental side could 
 explain this issue much better.  
 The quantum lower limit is given by $\tau_a$ at $F=F_a, (\delta_z=0)$,  
 where saturation is reached in $\tau_{T,d}$ for the adiabatic tunneling 
 time, which explains the Hartman effect in quantum tunneling \cite{Winful:2003}.
 \section{Concluding remarks}\label{sec:cr}
 For more insight and a conclusive judgment to the issue, we discuss 
 some points further in the following. 
 As already noted, Klaiber et al. presented in \cite{Klaiber:2015}, 
 a result concerning the tunneling dynamics and the attoclock, by 
 considering the experimental data of Boge et al. \cite{Boge:2013}. 
 The Keldysh parameter in this work, $\gamma_K\sim 0.8-4.9$, is in 
 the same range of the nonadiabatic field calibration of Hofmann et al. 
 \cite{Hofmann:2019}. 
 Klaiber et al. argued that the electron absorbs an effective number of 
 photon $\tilde{n}$ followed by a static tunneling at higher energy 
 $E=-I_p+ \tilde{n}\, \omega_0$ what they called a rule of thumb for 
 the region $\gamma_K\lesssim 1$. 
 The energy gain is defined semi-classically, with the assumption that 
 $\tilde{n}{\omega_0} ={\delta\cal{E}}$, where ${\delta\cal{E}}$ is an 
 energy change during the under-the-barrier motion \cite{Klaiber:2015}.   
 The rule is supposed to shift the exit point from the quasi-static 
 exit point with $x_{e,qs}=I_p/F=x_{C}$ (Fig. \ref{Hofdata1}) to the 
 exit point $x_e=x_{e,qs}-\delta x$ ($x_C$ is by no means correct, see 
 sec. \ref{sec:ep}).
 The view of Klaiber et al. is shown in Fig. \ref{Hofdata5} by (orange)  
 dashed-dotted curve with the exit point $x_{e,\tilde{n}}$ (compare with 
 Fig. 1 of \cite{Klaiber:2015}). 
 According to Klaiber et al. \cite{Klaiber:2015}, in the "nonadiabatic"  
 regime, the electron gains energy in the course of the 
 under-the-barrier motion, the nonadiabatic corrections raise the 
 energy level, and the tunnel exit shifts closer to the atomic core. 
 In their work, they compared the emission angle of the most probable 
 trajectory with the experimental data of Boge et al. \cite{Boge:2013} 
 for He-atom. 
 
\begin{figure}[t]
\resizebox{8.5cm}{5.50cm}{\includegraphics{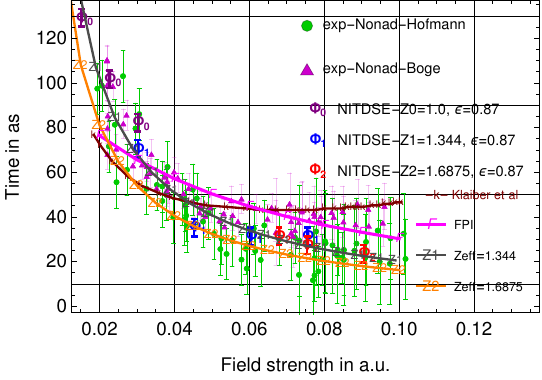}}
\vspace{-0.3cm}
\caption{\label{Hofdata7}\footnotesize (Color online) 
 The figure shows the time delay $\tau_{tion}$ given in eq \ref{dion} 
 for $Z_{eff}=1.6875$ and  
 $Z_{eff}=1.344$.   
 As in figs \ref{Hofdata2}, \ref{Hofdata3} the experimental 
 data of Hofmann \cite{Hofmann:2019} (green circles) are shown, 
 where we 
 included the nonadiabatic experimental data (purple triangle) 
 of Boge et al. \cite{Boge:2013} 
 and the result of Klaiber 
 et al. \cite{Klaiber:2015} (k, dark purple curve), see text.
 The NITDSE (see sec. \ref{sec:apdx} and \cite{IAIvanov:2014}), 
 as in Fig. \ref{Hofdata3}.}
\end{figure}

 To elucidate our discussion, we present Fig. \ref{Hofdata7}, where we 
 re-plot our result with the experimental data of Hofmann et al. and 
 include the experimental data of Boge et al. \cite{Boge:2013} and the 
 result of Klaiber et al. (see Fig. 2 of \cite{Klaiber:2015}). 
 We have to mention that both experimental results are from the same 
 group at ETH Zurich, where the recent experimental result of Hofmann 
 et al. \cite{Hofmann:2019} is supposed to be superior. 
 To compare with our result the data of Boge et al. and Klaiber et al. 
 have been converted from angle to time (in the same way as done by the   
 Hofmann data). 
 It is easy to see that the effective number of photons $\tilde{n}$ 
 assumed by Klaiber et al. have to be compared with our $\tilde{n}_F$.   
 However, it easily to find that $\tilde{n}<\tilde{n}_F\lesssim n_F=
 floor(\delta_z/\omega_0)$, see text before eq \ref{tion1} and Fig. 
 \ref{Hofdata5}.
 The tunneling step, which is supposed to occurs after multiphoton 
 absorption, is similar to that in our model. 
 However, the better agreement of our result with the experimental 
 data, as seen in Fig. \ref{Hofdata7}, confirms our approach and our 
 model as discussed in sec. \ref{sec:tt}, \ref{sec:dis}, \ref{sec:ir}.  
 The result of Klaiber et al. agrees well with the experimental data of 
 Boge et al. \cite{Boge:2013}, but its trend is not satisfactory.
 As already mentioned, the correct trend is determined (classically and 
 quantum-mechanically) by the $\sim \frac{1}{F}$ dependency of the time 
 delay, which can be inferred from the good agreement of our result 
 with the experimental data, compare Figs \ref{Hofdata2}, 
 \ref{Hofdata3}, \ref{Hofdata7}.
 
 Nevertheless, although Klaiber et al. \cite{Klaiber:2015} interpreting  
 the time delay differently \cite{Klaiber:2015}, we think that the 
 agreement with our result indicates that the emission angle or 
 equivalently, the time delay is caused by the barrier region 
 (under-the-barrier motion by \cite{Klaiber:2015}) in contrast to the 
 imaginary T-time picture, where the T-time is attributed to the tail 
 of the atomic potential \cite{Sainadh:2019}, although an equivalence 
 between the two pictures can be established in the adiabatic case, 
 as already discussed, and widely discussed in our previous works  
 \cite{Kullie:2015,Kullie:2016,Kullie:2018,Kullie:20181,Kullie:2020}.
 The comparison between the nonadiabatic and adiabatic case (of the 
 field calibration) in Fig. \ref{Hofdata6}, shows immediately that the 
 increased time delay in the latter case is due to the barrier itself, 
 the second term $\tau_{delt}$ in eq \ref{TdF}.     
 Strictly speaking, it is eliminated by Hofmann's nonadiabatic field 
 calibration.
 It affects a shift toward smaller field strength (increases the  
 self-interference contribution from the Winful point of view, eq 
 \ref{Winf}). 
 Equivalently, for the same F, a time delay contribution in the 
 adiabatic case emerges due to the second term $\tau_{delt}$, the 
 effects of the barrier itself.
 To conclude,  if we compare eq \ref{Tdi} with eqs  \ref{dion}, 
 \ref{Tsym} on one hand and the adiabatic with the nonadiabatic field 
 calibration on the other hand, and then both with each other, we see 
 that the field calibration maps $\Delta F $ to $\Delta t$ (or $\Delta 
 \theta$ of the streaking angle), which confirms the `real' T-time 
 picture (since $\delta_z$ is a real quantity). 
 For $F\rightarrow 0$, we reach a maximal time delay $\Delta t$, that 
 is $\lim\limits_{F\rightarrow 0}\tau_{dion} =\infty$, 
 $\lim\limits_{F\rightarrow 0} \delta_z=I_p, \tau_d\rightarrow \infty$. 
 It is in accordance with the measurement of a closed system, intrinsic 
 time  and the uncertainty principle as discussed in 
 \cite{Aharonov:2000} (for further discussion we kindly refer the 
 reader to our previous works \cite{Kullie:2015}, \cite{Kullie:2016} 
 and \cite{Kullie:2018}.)
 We think that our conclusion is relevant for the theory of tunneling 
 in general, especially that we found a relationship to the UTTP of 
 Winful (compare eq \ref{TdF} and eq \ref{Winf}), and to the Hartman 
 effect or the Hartman paradox \cite{Winful:2003}.      
 
 Finally, many experimental points 
 (compare Fig. \ref{Hofdata6}, \ref{Hofdata7}) are below the limit of 
 $\tau_{dion}$ ($\Delta \nu=0, \Delta n=0$ in eq \ref{tion1}, 
 \ref{tion}).
 This can not be explained this way. 
 Since the use of SAEA or the $Z_{eff}$ is not crucial as discussed in 
 \cite{Hofmann:2019}, it is difficult to understand this behavior.
 Multielectron effects are also small, they could be important for 
 small barrier width, but hardly explain this behavior for a larger barrier width.
 Although the experiment is challenging, improvement is desirable.   
 In particular, a refinement that reduces the spread of the 
 experimental point and improving the error bars is useful to 
 understand a tunneling contribution. 
\subparagraph*{Conclusion}
 In this work, we showed that our model is capable of describing the 
 experimental result of the nonadiabatic field calibration of Hofmann 
 et al. \cite{Hofmann:2019} for the attoclock, and we found a good 
 agreement with the experimental data. 
 Furthermore, we preformed calculations of the NITDSE for the $Z_{eff}$ 
 used in our model, which strongly support our obtained result and our 
 point of view.      
 Particularly, in our nonadiabatic picture, multiphoton absorption is 
 the most significant nonadiabatic effect. 
 The time it takes at a field strength $F\le F_a$, is a time delay with 
 respect to the ionization at atomic field strength $F_a$, where the BSI 
 sets up.
 The time delay generally consists of two terms.  
 The first term, $\tau_{dion}\,(\tau_{sym})$, is solely because of 
 $F<F_a$, and is it the time delay in the nonadiabatic field-ionization.
 The second term, $\tau_{delt}$, is a time delay due to the barrier 
 itself. 
 It represents a tunneling contribution; it is largest in the adiabatic 
 field calibration, for the maximum barrier height $\delta_z$, which 
 is discussed in previous work \cite{Kullie:2015}. 
 It saturates at the limit $F=F_a$, which explains the Hartman effect or 
 Hartman paradox.
 Our view is in accordance with the UTTP of Winful \cite{Winful:2003}.
 We also discussed the intermediate regime especially right below 
 the top of the barrier, where the tunneling probability is notably 
 high.

 With this, we think that we have made an important contribution to 
 resolving controversies related to the multiphoton and the tunneling 
 regimes, since Keldysh parameter $\gamma_K$ of eq \ref{gamK} is 
 usually not-strictly and vaguely applied in the strong-field regime. 
 Also the use of so-called classical exit point $x_C$, as usually done, 
 is by no mean a correct choice.  
 The Keldysh parameter indicates two limiting cases tunneling and 
 multiphoton regimes of the field-ionization, $\gamma_K\ll 1$, 
 $\gamma_K\gg 1$, respectively. 
 It is, however, mainly applied in the regime $\gamma_K \sim 1$
 for the strong-field-ionization and in attosecond science.  
   
 Even if one insists on two different interpretations of the 
 attoclock, nonadiabatic and adiabatic (which is important for the 
 tunneling theory), the time it takes in both cases is a time delay 
 with regard to the ionization at the atomic field strength $F_a$. 
 It is in accordance with the intrinsic dynamical time point of view  
 \cite{Busch:2008,Aharonov:2000,Kullie:2015,Kullie:2020}.
 Considering the experimental data of Hofman et al. \cite{Hofmann:2019} 
 to present the correct calibration (apart from the error bars), the 
 agreement presented in this work shows that after scattering with the 
 laser pulse the traversing of the barrier region  is essentially 
 driven by multiphoton ionization, in strong field interaction. 
 A tunneling contribution is possible, just below the top of the 
 barrier and can be associated with an intermediate regime or 
 intermediate tunneling, as discussed in sec. \ref{sec:ir}. 
 The attoclock receives a new boost, and the subtlety of the 
 experimental investigations is more demanding than ever before.
 The investigation of the tunneling or tunnel-ionization in the future 
 is essential to solve some of the questions regarding the 
 tunneling process itself. 
 The tunneling versus multiphoton ionization in the strong field, 
 the attosecond, and the ultrafast science have become more challenging 
 than ever. 
\appendix
\section{}
\subsection{Numerical Integration of the Time-Dependent Schr\"odinger 
Equation}\label{sec:apdx}

We follow closely the numerical procedure we used to solve the TDSE 
in \cite{IAIvanov:2014,IAIvanov_KTKIM:2016}.
We solve the TDSE for a single-electron atom with an effective central 
potential:
$\displaystyle V(r)=-\frac{Z_{eff}}{r}$ in the presence of a laser pulse: 
\begin{equation}
i \frac{\partial \Psi(\boldsymbol r)} {\partial t}=
\left(\hat H_{\rm atom} + \hat H_{\rm int}(t)\right)
\Psi(\boldsymbol r) \ .
\label{tdse1}
\end{equation}
We use velocity form for the operator $\hat H_{\rm int}(t)$
describing interaction of the atom with the laser field:
\begin{equation}
\hat H_{\rm int}(t) = {\boldsymbol A}(t)\cdot \hat{\boldsymbol p}\ ,
\label{gauge}
\end{equation}
where $\displaystyle \boldsymbol A(t)=-\int\limits_0^t 
\boldsymbol E(\tau)\ d\tau$ is the vector potential of the laser pulse, 
which for the geometry we employ (with quantization axis and pulse 
propagation direction along the $z$-axis), is defined as follows:
\begin{eqnarray}
A_x(t)&=& -{\frac{f(t)}{\omega \sqrt{1+\epsilon^2}}} F_0\cos{\omega t} \ , \nonumber \\
A_y(t)&=& {\frac{f(t)\epsilon}{\omega \sqrt{1+\epsilon^2}}} 
F_0\sin{\omega t} \ ,
\label{ef}
\end{eqnarray}
where $\epsilon=0.87$ is ellipticity of the pulse, $F_0$ its field 
strength (not to be confused with the {\it peak field strength} $F$ 
which we use in the formulas in the main text). The function $f(t)$ 
in eq \ref{ef} is 
the pulse envelope which we chose as: $\displaystyle f(t)= \sin^{16}(\pi t/ T_1)$, 
where $T_1=2T$, with  $T=2\pi/\omega$- an optical cycle corresponding 
to the fundamental frequency $\omega=0.062$ a.u., is a total duration 
of the pulse.
\begin{figure}[t]
\vspace{-5.50cm}
\resizebox*{6cm}{!}{\includegraphics{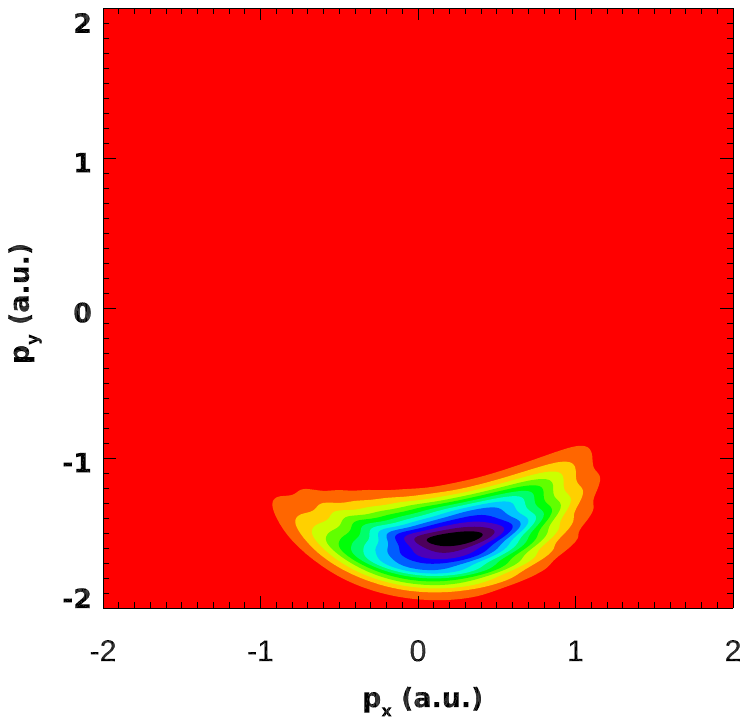}}
\end{figure}
\begin{figure}[t]
\vspace{-18.50cm}
\resizebox*{16cm}{!}{\includegraphics{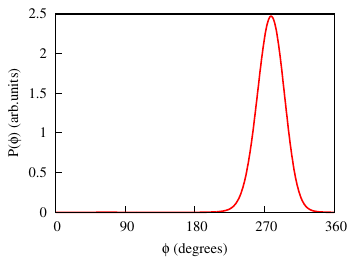}}
\vspace{-2.0cm}
\caption{\label{tdse}\footnotesize Photo-electron momentum distribution in the
polarization plane (above), and radially integrated distribution define by
Eq \ref{rint} (below). Field and target parameters: $F_0=0.12$ a.u.,
$Z_{eff}=1.6875$.}
\end{figure}
The initial state of the system is the ground $1s$ state of an atom with 
effective potential $\displaystyle V(r)=-\frac{Z_{eff}}{r}$. 
Solution of the TDSE is represented as a  series in spherical harmonics:
\begin{equation}
\Psi({\boldsymbol r},t)=
\sum\limits_{l,m} f_{l}(r,t) Y_l^m(\theta,\phi) \ ,
\label{basis}
\end{equation}
where spherical harmonics with orders up to $L_{\rm max}=100$ were used 
for the highest field strength $F_0=0.12$ a.u. we employed in the 
calculations.
The radial variable is treated by discretizing the TDSE on a grid with 
the step-size $\delta r=0.1$ a.u. in a box of the size $R_{\rm max}=400$ 
a.u. Necessary checks were performed to ensure that for these values 
of the parameters $L_{\rm max}$ and $R_{\rm max}$ convergence of the 
calculations has been achieved. The wave-function 
$\Psi({\boldsymbol r},t)$ was propagated in time using the matrix 
iteration method \cite{Nurhuda:1999}.

Ionization amplitude into a photo-electron state with asymptotic 
momentum $\boldsymbol p$ is computed by projecting the solution of the 
TDSE $\Psi({\boldsymbol r},T_1)$ at the end of the laser pulse on the 
scattering states $\phi_{\boldsymbol p}^-$ with ingoing boundary conditions. 

We are interested in photo-electron momenta distribution $P(p_x,p_y,0)$ 
in the polarization $(p_x,p_y)$-plane.
A typical distribution we obtain using the procedure we described above 
is shown in Fig \ref{tdse} (left) for $F_0=0.12$ a.u. and $Z_{eff}=1.6875$.  
An observable we are after is the offset angle, which for the pulse 
defined by eq \ref{ef} is the angle between the negative $y-$ 
direction and the ray pointing at the maximum of the photo-electron 
momentum distribution. 
To extract the offset angle, we follow the strategy we employed in 
\cite{IAIvanov:2014}). 
We compute the radially integrated distribution $P(\phi)$ defined as:
\begin{equation}
P(\phi)= \int\limits_0^{\infty} P(p_x,p_y,0)p\ dp \ ,
\label{rint}
\end{equation}
where $p=\sqrt{p_x^2+p_y^2}$, factor $p$ under the integral sign in 
eq \ref{rint} appears because of the area element in the 
$(p_x,p_y)$-plane, and angle $\phi$ is measured from the positive 
$x-$ direction. 

An example of the radially integrated distribution $P(\phi)$ is shown
in Fig \ref{tdse} (below). 
Offset angle now is determined as the location of the maximum of 
$P(\phi)$ minus 270 degrees.
\subparagraph*{Acknowledgments}
 O. Kullie would like to thank C. Hofmann for sending the experimental 
 data and FPI data shown in the figures, and  R. Boge for forwarding 
 the data presented in Fig 7, which was sent in the past when  
 a previous work \cite{Kullie:2018} was published.
 O. Kullie  would like to thank Prof. Martin Garcia from the 
 Theoretical Physics of the Institute of Physics at the University 
 of Kassel for his kind support. 
 I. Ivanov acknowledges support by the Institute for Basic Science
under the grant number IBS-R012-D1.
%
\end{document}